# LIST DECODING TENSOR PRODUCTS AND INTERLEAVED CODES

PARIKSHIT GOPALAN, VENKATESAN GURUSWAMI, AND PRASAD RAGHAVENDRA

ABSTRACT. We design the first efficient algorithms and prove new combinatorial bounds for list decoding tensor products of codes and interleaved codes.

- We show that for *every* code, the ratio of its list decoding radius to its minimum distance stays unchanged under the tensor product operation (rather than squaring, as one might expect). This gives the first efficient list decoders and new combinatorial bounds for some natural codes including multivariate polynomials where the degree in each variable is bounded.

- We show that for *every* code, its list decoding radius remains unchanged under $m$-wise interleaving for an integer $m$. This generalizes a recent result of Dinur *et al.* [6], who proved such a result for interleaved Hadamard codes (equivalently, linear transformations).

- Using the notion of generalized Hamming weights, we give better list size bounds for *both* tensoring and interleaving of binary linear codes. By analyzing the weight distribution of these codes, we reduce the task of bounding the list size to bounding the number of close-by low-rank codewords. For decoding linear transformations, using rank-reduction together with other ideas, we obtain list size bounds that are tight over small fields.

Our results give better bounds on the list decoding radius than what is obtained from the Johnson bound, and yield rather general families of codes decodable beyond the Johnson bound.

## 1. INTRODUCTION

The decoding problem for error-correcting codes consists of finding the original message given a corrupted version of the codeword encoding it. When the number of errors in the codeword is too large, and in particular could exceed half the minimum distance of the code, unambiguous recovery of the original codeword is no longer always possible. The notion of *list decoding* of error-correcting codes, introduced by Elias [10] and Wozencraft [35], provides one avenue for error-correction in such high noise regimes. The goal of a list decoding algorithm is to efficiently recover a list of all codewords within a specified Hamming radius of an input string. The central problem of list decoding is to identify the radius up to which this goal is tractable, both combinatorially (in terms of the output list being guaranteed to be small, regardless of the input) and algorithmically (in terms of being able to find the list efficiently).

The classical Johnson bound shows that at least combinatorially, list decoding always allows one to correct errors beyond half the minimum distance. It states that every code of distance $\delta$ over $\mathbb{F}_q$ is list-decodable up to the Johnson radius $J_q(\delta)$ which lies in the range $(\delta/2, \delta)$. However, the Johnson bound is oblivious to the structure of the code; it only depends on its

Research was supported in part by NSF CCF-0343672 and a David and Lucile Packard Fellowship.

minimum distance. Potentially, a code might be list-decodable at larger error-radii than what is guaranteed by the Johnson bound. The question of identifying the precise radius up to which list decoding is tractable for a family of codes is a challenging problem. Despite much progress in designing list decoding algorithms over the last decade, this problem is still open even for well-studied codes such as Reed-Solomon and Reed-Muller codes.

On the algorithmic side, following the breakthrough results of Goldreich-Levin [11] and Sudan [28] which gave list decoders for Hadamard codes and Reed-Solomon codes respectively, there has been tremendous progress in devising list decoders for various codes (see the surveys [15, 16, 29]). This study has had substantial impact on other areas such as complexity theory [30, 32], cryptography [11, 1] and computational learning [19, 21]. Examples of codes which are known to be list-decodable beyond the Johnson bound have been rare: Extractor codes [31, 14], folded Reed-Solomon codes [24, 17], group homomorphisms [6] and Reed-Muller codes over small fields [12] are the few examples known to us.

A natural way to design new error-correcting codes from old ones is via various *product* operations on these codes. In this work, we study the effect of two basic product operations, tensoring and interleaving, on list-decodability. In what follows, $[q]$ stands for an alphabet of size $q$, for example $\{0, 1, \ldots, q-1\}$.

**Definition 1.1.** *(Tensor Product)* Given two linear codes $\mathcal{C}_1 \subseteq [q]^{n_1}$ and $\mathcal{C}_2 \subseteq [q]^{n_2}$, their tensor product $\mathcal{C}_2 \otimes \mathcal{C}_1$ consists of all matrices in $[q]^{n_2 \times n_1}$ whose rows belong to $\mathcal{C}_1$ and columns belong to $\mathcal{C}_2$. For a code $\mathcal{C} \subseteq [q]^n$, its $m$-wise tensor product for $m \geqslant 1$ is a code of length $n^m$ defined inductively as $\mathcal{C}^{\otimes 1} = \mathcal{C}$ and $\mathcal{C}^{\otimes m} = \mathcal{C}^{\otimes (m-1)} \otimes \mathcal{C}$ for $m > 1$.

For example, Reed-Muller codes in $m$ variables where the degree in each variable is restricted to $d$ can be viewed as the $m$-wise tensor of Reed-Solomon codes. Our algorithm does not require the $\mathcal{C}_i$s to be linear, but we make the assumption since the tensor of two non-linear codes might be empty. Using $\delta(\mathcal{C})$ and $R(\mathcal{C})$ to denote the normalized distance and rate of $\mathcal{C}$ respectively, it follows that $\delta(\mathcal{C}^{\otimes m}) = \delta(\mathcal{C})^m$ and $R(\mathcal{C}^{\otimes m}) = R(\mathcal{C})^m$. Hence for tensor products, we are primarily interested in the setting where $m$ is either constant or a slowly growing function of the block length.

**Definition 1.2.** *(Interleaved Codes)* The $m$-wise *Interleaving* (or interleaved product) $\mathcal{C}^{\odot m}$ of the code $\mathcal{C} \subseteq [q]^n$ consists of $n \times m$ matrices over $[q]$ whose columns are codewords in $\mathcal{C}$. Each row is treated as a single symbol, thus $\mathcal{C}^{\odot m} \subseteq [q^m]^n$.

In other words, under $m$-wise interleaving, $m$ independent messages are encoded using $\mathcal{C}$, and the $m$ symbols in each position are juxtaposed together into a single symbol over a larger alphabet. For instance, linear transformations from $\mathbb{F}_2^n \to \mathbb{F}_2^n$ can be viewed as interleaved Hadamard codes. It is easy to see that $\delta(\mathcal{C}^{\odot m}) = \delta(\mathcal{C})$, $R(\mathcal{C}^{\odot m}) = R(\mathcal{C})$ but the alphabet grows from $[q]$ to $m$-dimensional vectors over $[q]$. So unlike for tensors, for interleaving $m$ could be as large as polynomial in the block length; indeed our results hold for any $m$.

1.1. **Tensor products and interleaved codes: Prior work and motivation.**

1.1.1. *Tensoring.* Tensor products occupy a central place in coding theory, so much so that tensor product codes are typically referred to as just product codes. Product codes provide a convenient way to construct longer codes from shorter component codes. Elias [9] used tensor product of Hamming codes to construct the first explicit codes with positive rate for communication over a binary symmetric channel. The structure of product codes enables decoding them along columns and rows, and the column decoder can provide valuable reliability information to the row decoder [26]. Product codes find many uses in practice; for example, the product of two Reed-Solomon codes is used for encoding data on DVDs.

More recently, tensor products have found applications in several areas of theoretical CS such as hardness of approximation [8, 20] and constructions of locally testable codes [22]. The effect of tensoring on on local testability of codes has been extensively studied [2, 7, 34, 4]. Tensor products admit a natural tester (which checks a random row/column) that has a certain "robustness" property [2]. Exploiting this, by recursive tensoring one can obtain simple constructions of locally testable codes with non-trivial parameters, starting from *any* reasonable constant-sized code. But to our knowledge, there seems to no prior work focusing on the effect of tensoring on the list decoding radius. In particular, the best combinatorial bound known for the list decoding radius seems to have been the Johnson bound, and we are unaware of an efficient algorithm that decodes $\mathcal{C}_2 \otimes \mathcal{C}_1$ up to the Johnson bound, assuming that such algorithms exists for each $\mathcal{C}_i$. A sufficiently strong result on list decoding tensor products could lead to a simple, recursive construction of list-decodable codes starting from any small code of good distance. Our results are optimal in terms of the radius to which we decode, but not strong enough in terms of output list-size guarantee to obtain such a result.

1.1.2. *Interleaving.* At a high level, interleaving is a way to arrange data in a non-contiguous way in order to increase performance. Interleaving is used in practical coding systems to group together the symbols of several codewords as a way to guard against *burst* errors. A burst error could cause too many errors within one codeword, making it unrecoverable. However, with interleaving, the errors get distributed into a small, correctable number of errors in each of the interleaved codewords. This is quite important in practice; for example the data in a CD is protected using cross-interleaved Reed-Solomon coding (CIRC). Though code interleaving has been implicitly studied for its practical importance, our work appears to be the first to study it in generality as a formal product operation. We describe some recent theoretical work on interleaved codes which set the stage for our investigation.

The problem of decoding interleaved Reed-Solomon codes from a large number of *random* errors was tackled in [5, 3]. The folded Reed-Solomon codes constructed by Guruswami and Rudra [17] (which achieve list decoding capacity), and their precursor Parvaresh-Vardy codes [24], are both sub codes of interleaved Reed-Solomon codes, where the $m$ interleaved codewords are carefully chosen to have dependencies on each other. Dinur *et al.* in their work on list decoding group homomorphisms studied interleaved Hadamard codes, which are essentially all linear transformations from $\mathbb{F}_q^n \to \mathbb{F}_q^m$ [6]. Their work raised the question of how the interleaving operation affects the list decoding radius of arbitrary codes, and motivated our results.

1.2. **Brief high-level description of key results.** We start with some terminology. The distance $\text{dist}(\mathcal{C})$ of a code $\mathcal{C} \subseteq [q]^n$ is the minimum Hamming distance between two distinct codewords in $\mathcal{C}$, and its *relative distance* is defined as $\text{dist}(\mathcal{C})/n$. For a code $\mathcal{C}$ of block length $n$ and $0 < \eta < 1$, the *list-size for radius $\eta$*, denoted $\ell(\mathcal{C}, \eta)$, is defined as the maximum number of codewords in a Hamming ball of radius $\eta m$. Informally, the *list decoding radius* (LDR) of $\mathcal{C}$ is the largest $\eta$ such that for every constant $\varepsilon > 0$, the list size for radius $(\eta - \varepsilon)$ is bounded by some function $f(q, \varepsilon)$ independent of the block length $n$.[1]

Our main result on tensor products is the following: if $\mathcal{C}$ has relative distance $\delta$ and LDR $\eta$, then the list decoding radius of the $m$-wise product $\mathcal{C}^{\otimes m}$ is $\eta\delta^{m-1}$. In other words, the ratio of LDR to relative distance is preserved under tensoring.

For interleaved codes, we prove that the LDR remains unchanged irrespective of the number of interleaves. In particular, if $\mathcal{C}$ has relative distance $\delta$, and $\mathcal{C}^{\odot m}$ is its $m$-wise interleaving, then for every $\eta < \delta$ one has $\ell(\mathcal{C}^{\odot m}, \eta) \leqslant A \cdot \ell(\mathcal{C}, \eta)^B$ where $A, B$ are constants depending only on $\delta, \eta$ and *independent of $m$*.

**Organization.** Formal statement of all the results of this work appear in Section 2. These include the results described above which apply to arbitrary codes, along with improved list size bounds for special cases like binary linear codes and transformations. We present our bounds for interleaved codes in Section 3, and our decoding algorithm for tensor products together with a sketch of its analysis in Section 4. The formal analysis of the tensor decoding algorithm appears in Section 5. In Section 6, we use the notion of generalized Hamming weights to derive improved list-size bounds for tensor products and interleavings of binary linear codes. Finally, in Section 7, we show list size bounds for linear transformations that are tight over small fields.

## 2. Our Results

### 2.1. List decoding tensor products.
Given two codes $\mathcal{C}_1 \subseteq [q]^{n_1}$ and $\mathcal{C}_2 \subseteq [q]^{n_2}$, their tensor product $\mathcal{C}_2 \otimes \mathcal{C}_1$ consists of all matrices in $[q]^{n_2 \times n_1}$ whose rows belong to $\mathcal{C}_1$ and columns belong to $\mathcal{C}_2$.

We design a generic algorithm that list decodes $\mathcal{C}_2 \otimes \mathcal{C}_1$ using list decoders for $\mathcal{C}_1$ and $\mathcal{C}_2$ as subroutines, and bound the list-decoding radius by analyzing its performance. Further, if we have efficient list decoders for $\mathcal{C}_1$ and $\mathcal{C}_2$, then we get an efficient algorithm for list decoding $\mathcal{C}_2 \otimes \mathcal{C}_1$. A brief overview of this algorithm is given below in Section 2.1.2. Our main result on list decoding tensor products is the following.

**Theorem 2.1.** *Let $\mathcal{C}_1 \subseteq [q]^{n_1}$ and $\mathcal{C}_2 \subseteq [q]^{n_2}$ be codes of relative distance $\delta_1$ and $\delta_2$ respectively, and $0 < \eta_1, \eta_2 < 1$. Define $\eta^* = \min(\delta_1\eta_2, \delta_2\eta_1)$. Then*

$$\ell(\mathcal{C}_2 \otimes \mathcal{C}_1, \eta^* - \varepsilon) \leqslant q^{O_{\varepsilon, \delta_1, \delta_2}(\log \ell(\mathcal{C}_1, \eta_1) \log \ell(\mathcal{C}_2, \eta_2))} .$$

*In particular, if the LDR of $\mathcal{C}_1, \mathcal{C}_2$ are $\eta_1, \eta_2$ respectively, then the LDR of $\mathcal{C}_2 \otimes \mathcal{C}_1$ is at least $\eta^*$.*

The decoding radius achieved in Theorem 2.1 is in fact tight: it is easily shown that assuming that $\mathcal{C}$ cannot be list decoded beyond radius $\eta$, $\mathcal{C}^{\otimes 2}$ cannot be list decoded beyond $\delta\eta$ (Lemma 4.1). As a corollary, the ratio of the list decoding radius to the relative distance stays unchanged under repeated tensoring.

---

[1] Here it is implicitly implied that the codes $\mathcal{C}$ belong to an infinite family of codes of increasing block length.

**Corollary 2.2.** *If $\mathcal{C}$ has relative distance $\delta$ and list decoding radius $\eta$, then the list decoding of the $m$-wise tensor product $\mathcal{C}^{\otimes m}$ is $\eta\delta^{m-1}$.*

The bounds that we get on list-size for $\mathcal{C}^{\otimes m}$ are doubly exponential in $m$. Improving this bound to singly exponential in $m$, say $\exp(O(m))$, could have interesting applications, such as a simple construction of list-decodable codes by repeated tensoring, with parameters strong enough for the many complexity-theoretic applications which currently rely crucially on Reed-Solomon list decoding. We are able to obtain some improvements for the case of tensoring binary linear codes; this is described in Theorem 2.8.

**Comparison to Johnson bound.** Even assuming that each of $\mathcal{C}_1$ and $\mathcal{C}_2$ are decodable only up to the Johnson bound, Theorem 2.1 gives a bound that is significantly better than the Johnson bound, since by convexity $J_q(\delta_1\delta_2) < \min(\delta_1 J_q(\delta_2), \delta_2 J_q(\delta_1)) \leqslant \eta^*$.

2.1.1. *Implications for natural codes.* Theorem 2.1 gives new bounds on list decoding radius for some natural families of codes, which we discuss below.

**Reed-Solomon tensors.** Let $\mathrm{RS}[n,k]_q$ denote the Reed Solomon code consisting of evaluations of degree $k$ polynomials over a set $S \subseteq \mathbb{F}_q$ of size $n$, with distance $\delta = 1 - k/n$. Such codes are list-decodable up to the Johnson radius $J(\delta) = 1 - \sqrt{1-\delta}$ using the Guruswami-Sudan algorithm [18]. The $m$-wise tensor product of such a RS code is a $[n^m, k^m]_q$ code consisting of evaluations on $S^m \subseteq \mathbb{F}_q^m$, of multivariate polynomials in $m$ variables with individual degree of each variable being at most $k$. Parvaresh *et al.* considered the problem of decoding these product codes, and extended the Reed-Solomon list decoder to this setting [23]. This yields relatively weak bounds, and as they note, by reducing the problem to decoding Reed-Muller codes of order $mk$, one can do better [25]. Still, these bounds are weaker than the Johnson radius $J(\delta^m)$ of the $m$-wise product, and in fact become trivial when $m > n/k$. Our results give much stronger bounds, and enable decoding *beyond* the Johnson bound.

**Corollary 2.3.** *The $m$-wise tensor of the $\mathrm{RS}[n,k]_q$ Reed-Solomon code is efficiently list-decodable up to a radius $\delta^{m-1}J(\delta) - \varepsilon$, where $\delta = 1 - k/n$ is the relative distance of $\mathrm{RS}[n,k]_q$, and $\varepsilon > 0$ is arbitrary.*

One can compare this with the Johnson bound $J(\delta^m)$ by noting that

$$\delta^{m-1}J(\delta) = \delta^m\left(\frac{1}{2} + \frac{1}{4}\delta + \frac{3}{8}\delta^2 + \cdots\right), \quad J(\delta^m) = \delta^m\left(\frac{1}{2} + \frac{1}{4}\delta^m + \frac{3}{8}\delta^{2m} + \cdots\right).$$

**Hadamard tensors.** Let Had be the $[q^k, k]_q$ Hadamard code, where $a \in \mathbb{F}_q^k$ is encoded as the vector $\{a \cdot x\}_{x \in \mathbb{F}_q^k}$. $\mathcal{C}$ is list-decodable up to its relative distance $\delta = 1 - 1/q$. The $m$-wise tensor product $\mathcal{C}^{\otimes m}$ consists of all "block-linear" polynomials. Specifically, each codeword in $\mathcal{C}^{\otimes m}$ is a polynomial $P$ on $m \times k$ variables given by $x = (x^{(1)}, x^{(2)}, \ldots, x^{(m)})$ where each $x^{(i)} = (x_1^{(i)}, \ldots, x_k^{(i)})$, such that for each $i$, $P$ is a linear function in $x^{(i)}$ for each fixing of the other variables.

**Corollary 2.4.** *The $m$-wise tensor of $[q^k, k]_q$ Hadamard codes has list decoding radius equal to its relative distance $\left(1 - \frac{1}{q}\right)^m$.*

This result is interesting in light of a conjecture by Gopalan *et al.* stating that Reed-Muller codes of degree $m$ over $\mathbb{F}_q$ are list-decodable up to the minimum distance (they proved this result for $q = 2$) [12]. Our result shows $m$-wise Hadamard tensors which are a natural sub code of order $m$ Reed-Muller codes (with better distance but lower rate) are indeed list-decodable up to the minimum distance.

2.1.2. *Tensor decoder overview.* Our algorithm for list decoding $\mathcal{C}_2 \otimes \mathcal{C}_1$ starts by picking small random subsets $S \subset [n_2]$ and $T \subset [n_1]$ of the rows and columns respectively. We assume that we are given the codeword restricted to $S \times T$ as advice. By alternately running the row and column decoders, we improve the quality of the advice. We show that after four alternations, one can recover the codeword correctly with high probability (over the choice of $S$ and $T$). An obstacle in decoding tensor product codes is that some of the rows/columns could have every high error-rates, and decoding those rows/columns of the received word gives incorrect codewords. We show that the advice string allows us to identify such rows/columns with good probability, thus reducing the problem to decoding from (few) errors and (many) erasures. The scheme of starting with a small advice string and recovering the codeword via a series of self-correction steps has been used for list decoding Hadamard codes and Reed-Muller codes. Our work is the first (to our knowledge) that applies it outside the setting of algebraic codes defined over a vector space.

2.2. INTERLEAVED CODES. Armed with a list decoding algorithm for $\mathcal{C}$, a naive attempt at list decoding $\mathcal{C}^{\odot m}$ would proceed as follows: List decode each column of the received word separately to obtain $m$ different lists $\{L_1, \ldots, L_m\}$, then iterate over all matrices with first column from $L_1$, second column from $L_2$, etc., and output those close enough to the received word. The naive algorithm described above yields the following simple product bound

$$(2.1) \qquad \ell(\mathcal{C}, \eta) \leqslant \ell(\mathcal{C}^{\odot m}, \eta) \leqslant \ell(\mathcal{C}, \eta)^m$$

This upper bound is unsatisfactory since even if $\ell(\mathcal{C}, \eta) = 2$, the upper bound on $\ell(\mathcal{C}^{\odot m}, \eta)$ is $2^m$. Recent work of Dinur *et al.* [6] overcame this naive product bound when the codes being interleaved arise from group homomorphisms. To this end, they extensively used properties of certain set systems that arise in the context of group homomorphisms.

Surprisingly, we show that the product bound can be substantially improved for *every* code $\mathcal{C}$. In fact, the list size bound we obtain is independent of the number of interleavings $m$ (as in the above-mentioned results of Dinur *et al.* [6] ).

**Theorem 2.5.** *Let $\mathcal{C}$ be a code of relative distance and let $\eta < \delta$. Define $b = \lceil \frac{\eta}{\delta - \eta} \rceil$ and $r = \lceil \log \frac{\delta}{\delta - \eta} \rceil$. Then, for all integers $m \geqslant 1$, we have*

$$(2.2) \qquad \ell(\mathcal{C}^{\odot m}, \eta) \leqslant \binom{b + r}{r} \ell(\mathcal{C}, \eta)^r \ .$$

The implies that if $\mathcal{C}$ is list-decodable up to radius $\eta$, then so is $\mathcal{C}^{\odot m}$. The condition $\eta < \delta$ in Theorem 2.5 is necessary, as it is easily shown that $\ell(\mathcal{C}^{\odot m}, \delta) \geqslant 2^m$ (unless $\mathcal{C}$ is trivial and has only one codeword).

2.2.1. *Proof technique.* The proof of Theorem 2.5 relies on a simple observation which we outline below. Assume that we list decode the received word corresponding to the first column, to get a list of candidate codewords for that column and pick one codeword from this list. Rows where the first column of the received word differs from this codeword correspond to errors, hence we can replace those rows by erasures. Thus for the second column, some of the error locations are erased, which makes the decoding easier. Of course, if the codeword is close to the received word, then there may be very few (or no) erasures introduced. But we show there are only a few codewords in the list that are very close to the received word. Extending this intuition, we construct a tree of possible codewords for each column and show that the tree is either shallow or it does not branch too much.

2.3. BETTER LIST-SIZE BOUNDS USING GENERALIZED HAMMING WEIGHTS. For the case of binary linear codes, we are able to improve the list size upper bounds for both tensoring and interleaving (Theorems 2.1 and 2.5 above) using a common technique. We now describe the underlying idea and the results it yields.

2.3.1. *Method overview.* Codewords of both interleaved and 2-wise tensor products are naturally viewed as matrices. We bring the rank of these matrices into play, and argue that if the rank of a codeword is large, then its Hamming weight is substantially higher than the distance of the code. It turns out that this phenomenon is captured exactly by a well-studied notion in coding theory called *generalized Hamming weights* (see the survey [33]) that is also closely related to list decoding from erasures [13]. The precise connection is that if a codeword of $\mathcal{C}^{\odot m}$ has rank $r$, then its relative Hamming weight is at least the $r^{th}$ generalized Hamming weight $\delta_r(\mathcal{C})$ of $\mathcal{C}$. Similarly, rank $r$ codewords in $\mathcal{C} \otimes \mathcal{C}$ have weight at least $\delta_r(\mathcal{C})\delta(\mathcal{C})$.

For binary codes, for $r$ large enough, $\delta_r(\mathcal{C})$ approaches $2\delta$. The Johnson radius of $2\delta$ exceeds $\delta$. Therefore, for $r = r(\delta, \eta)$ large enough, the number of codewords in a Hamming ball of radius $\eta < \delta$ whose pairwise differences all have rank $> r$ can be bounded from above using the Johnson bound. Using the deletion argument from [12], the task of bounding the list-size for radius $\eta$ now reduces to bounding the number of rank $\leqslant r$ codewords within radius $\eta$. We accomplish this task, for both interleaved and product codes, using additional combinatorial ideas. We remark that our use of the deletion argument is more sophisticated than in [12], since most of the work goes into bounding the list-size for the low-rank case.

We note that the reason the above approach does not work for non-binary alphabets is that the generalized Hamming weight $\delta_r(C)$ may not be larger than $\frac{q}{q-1}\delta$ for $q$-ary codes.

2.3.2. *Results for interleaved codes.* Theorem 2.5 showed that for any code $\mathcal{C}$ of distance $\delta$ and for any $\eta < \delta$, the list-size for the $m$-wise interleaved code $\mathcal{C}^{\odot m}$ is bounded by $\ell(\mathcal{C}, \eta)^{\lceil \log \frac{\delta}{\delta - \eta} \rceil}$. Note that for $\eta \to \delta$, the exponent grows without bounds. For binary linear codes, using the above generalized Hamming weights based approach, we can improve this bound to a fixed polynomial in $\ell(\mathcal{C}, \eta)$, removing the dependence on $\log(1/(\delta - \eta))$ in the exponent.

**Theorem 2.6.** *For any binary linear code with distance $\delta$, we have:*

$$(2.3) \qquad \ell(\mathcal{C}^{\odot m}, \eta) \leqslant \frac{4}{\delta^4} \left( \frac{2\ell(\mathcal{C}, \eta)}{\delta^2(\delta - \eta)} \right)^{\lceil \log \frac{2}{\delta^2} \rceil}.$$

Given a binary linear error-correcting code $\mathcal{C}$, the Johnson bound states that the list size at radius $J_2(\delta) - \varepsilon$ is bounded by $O(\varepsilon^{-2})$. We can show that essentially the same list-size bound holds for $\mathcal{C}^{\odot m}$, provided the distance $\delta$ is bounded away from $\frac{1}{2}$.

**Theorem 2.7.** *For every $\delta < \frac{1}{2}$, there exists a constant $c_\delta$ such that for every binary linear code $\mathcal{C}$ of relative distance $\delta$,*
$$\ell(\mathcal{C}^{\odot m}, J_2(\delta) - \varepsilon) \leqslant c_\delta \varepsilon^{-2}.$$

2.3.3. *Result for tensor product.* Applying the above approach to the tensor product of two codes, we prove the following (Theorem 6.15). Note that the list size is at most a fixed polynomial in the list sizes $\ell(\mathcal{C}_1, \eta_1)$ and $\ell(\mathcal{C}_2, \eta_2)$ of the original codes, instead of the quasipolynomial dependence in Theorem 2.1.

**Theorem 2.8.** *Suppose $\mathcal{C}_1 \subseteq \mathbb{F}_2^{n_1}$ and $\mathcal{C}_2 \subseteq \mathbb{F}_2^{n_2}$ are binary linear codes of relative distance $\delta_1$ and $\delta_2$ respectively. Let $\eta_1 \leqslant \delta_1$ and $\eta_2 \leqslant \delta_2$. Define $\eta^* = \min(\delta_1\eta_2, \delta_2\eta_1)$ and $r = \lceil \log(\frac{2}{\delta_1\delta_2}) \rceil$. Then*
$$\ell(\mathcal{C}_2 \otimes \mathcal{C}_1, \eta^* - \varepsilon) \leqslant 2^{O(r^2)} \ell(\mathcal{C}_1, \eta_1)^r \ell(\mathcal{C}_2, \eta_2)^r \varepsilon^{-2r} .$$

2.4. **List decoding linear transformations.** Let $\mathrm{Lin}(\mathbb{F}_q, k, m)$ denote the set all linear transformations $L : \mathbb{F}_q^k \to \mathbb{F}_q^m$. The code $\mathrm{Lin}(\mathbb{F}_q, k, m)$ is nothing but the $m$-wise interleaving $\mathrm{Had}^{\odot m}(\mathbb{F}_q, k)$ of the Hadamard code $\mathrm{Had}(\mathbb{F}_q, k)$. Let $\ell(\mathrm{Lin}(\mathbb{F}_q, k, m), \eta)$ denote the maximum list size for the code $\mathrm{Lin}(\mathbb{F}_q, k, m)$ at a distance $\eta$. Dinur *et al.* [6] show that $\ell(\mathrm{Lin}(\mathbb{F}_2, k, m), 1/2 - \varepsilon) \leqslant O(\varepsilon^{-4})$ and $\ell(\mathrm{Lin}(\mathbb{F}_q, k, m), 1 - 1/q - \varepsilon) \leqslant O(\varepsilon^{-C})$ for some constant $C$ for general $q$. The best lower-bound known for any field is $\Omega(\varepsilon^{-2})$.

Being a general result for all codes, Theorem 2.5 only gives a quasipolynomial bound for the special case of linear transformations. By specializing the above generalized Hamming weights approach to the case of linear transformations, and using more sophisticated arguments based on decoding from erasures for the low-rank case, we prove the following stronger bounds for list decoding linear transformations over $\mathbb{F}_2$.

**Theorem 2.9.** *There is a constant $C$ such that for all positive integers $k, m$,*
$$\ell(\mathrm{Lin}(\mathbb{F}_2, k, m), \frac{1}{2} - \varepsilon) \leqslant C\varepsilon^{-2} .$$

For arbitrary fields $\mathbb{F}_q$, we prove the following bounds, the first is asymptotically tight for small fields while the second is independent of $q$ and improves on the bound of Dinur *et al.* .

**Theorem 2.10.** *There is an absolute constant $C'$ such that for every finite field $\mathbb{F}_q$,*
$$\ell(\mathrm{Lin}(\mathbb{F}_q, k, m), 1 - \frac{1}{q} - \varepsilon) \leqslant C' \min(q^6\varepsilon^{-2}, \varepsilon^{-5}) .$$

# 3. Interleaved Codes

In this section, $\mathcal{C} \subset [q]^n$ will be an arbitrary code (possibly non-linear) over an alphabet $[q]$. We will use $\ell(\eta)$ and $\ell^{\odot m}(\eta)$ for $\ell(\mathcal{C}, \eta)$ and $\ell(\mathcal{C}^{\odot m}, \eta)$ respectively. Let $d_q(c_1, c_2)$ denote the Hamming distance between strings in $[q]^n$ and $\Delta_q(c_1, c_2) = d_q(c_1, c_2)/n$ denote the normalized Hamming distance. We drop the subscript when the alphabet is clear from context. For $r \in [q]^n$, $\mathrm{B}(r, \eta) \subseteq [q]^n$ denotes the Hamming ball centered at $r$ of radius $\eta n$. We use $C$ for codewords of $\mathcal{C}^{\odot m}$ and $c$ for codewords of $\mathcal{C}$. We will interchangeably view $C$ as a matrix in $[q]^{n \times m}$ and a vector in $[q^m]^n$. For a $k \times m$ matrix $A$, $a_1, \ldots, a_m$ will denote its columns, $a[1], \ldots, a[k]$ will denote the rows, and $A_{\leqslant i}$ will denote the $k \times i$ matrix$(a_1, \ldots, a_i)$.

Given an algorithm DecodeC that can list decode $\mathcal{C}$ up to radius $\eta$, it is easy to give an algorithm DecodeC$^{\odot m}$ that uses DecodeC as a subroutine and runs in time polynomial in the list-size and $m$; we present this algorithm in Section 3.1. Thus it suffices to bound the list-size to prove Theorem 2.5. We do this by giving a (possibly inefficient) algorithm, which identifies rows where errors have occurred and erases them. Erasing a set $S \subset [n]$ of co-ordinates is equivalent to puncturing the code by removing those indices. Given $r \in [q]^n$, we use $r^{-S}$ to denote its projection on to the co-ordinates in $[n] \setminus S$.

**Definition 3.1.** *(Erasing a subset)* Given a code $\mathcal{C} \subseteq [q]^n$, erasing the indices corresponding to $S \subseteq [n]$ gives the code $\mathcal{C}^{-S} = \{c^{-S} : c \in \mathcal{C}\} \subseteq [q]^{n-|S|}$.

Let $|S| = \mu n$. We will only consider the case that $\mu < \delta$. It is easy to see that the resulting code $\mathcal{C}^{-S}$ has distance $d(\mathcal{C}^{-S}) \geqslant (\delta - \mu)n$. There is a 1-1 correspondence between codewords in $\mathcal{C}$ and their projections in $\mathcal{C}^{-S}$. For the code $\mathcal{C}^{-S}$, it will be convenient to consider standard Hamming distance, to avoid normalizing by $1 - \mu$. For $\eta < 1 - \mu$, let $\ell^{-S}(\eta)$ be the maximum number of codewords of $\mathcal{C}^{-S}$ that lie in a Hamming ball of radius $\eta n$ in $[q^m]^{n(1-\mu)}$.

**Lemma 3.2.** *For any $\eta < 1 - \mu$, $\ell^{-S}(\eta) \leqslant \ell(\eta + \mu)$.*

*Proof.* Take a received word $r^{-S} \in [q]^{n(1-\mu)}$ so that there are $L$ codewords $c_1^{-S}, \ldots, c_L^{-S}$ satisfying $d(r^{-S}, c_i^{-S}) \leqslant \eta n$. Define $r \in [q]^n$ by fixing values at the set $S$ arbitrarily. By the triangle inequality, $d(r, c_i) \leqslant (\eta + \mu)n$, showing that $\ell(\eta + \mu) \geqslant L$. $\qquad\square$

Assume we have a procedure List-Decode that takes as inputs set $S \subseteq [n]$, $r \in [q]^n$, an error parameter $e$ and returns all codewords $c \in C$ so that $d(c^{-S}, r^{-S}) \leqslant e$ (it need not be efficient). We use it to give an algorithm for list decoding $\mathcal{C}^{\odot m}$, which identifies rows where errors have occurred and erases them. Assume we have fixed $C_{\leqslant i} = (c_1, \ldots, c_i)$. We erase the set of positions $S$ where $C_{\leqslant i} \neq R_{\leqslant i}$ and then run a list decoder for $\mathcal{C}^{-S}$ on $r_{i+1}$. The crucial observation is that since the erased positions all correspond to errors, the number of errors drops by $|S|$. The distance might also drop by $|S|$, but since $\eta < \delta$ to begin with, this tradeoff works in our favor.

---

**Algorithm 1.** Erase-Decode
**Input:** $R \in [q^m]^n, \eta$.
**Output:** List $\mathcal{L}$ of all $C \in \mathcal{C}^{\odot m}$ so that $\Delta_{q^m}(R, C) \leqslant \eta$.

Set $S_1 = \phi, \mu_1 = 0$.
For $i = 1, \ldots, m$
1.  Set $\mathcal{L}_i = $ List-Decode$(S_i, r_i, (\eta - \mu_i)n)$.
2.  Choose $c_i \leftarrow \mathcal{L}_i$.
3.  Set $S_{i+1} = \{j \in [n] \ s.t. \ C_{\leqslant i}[j] \neq R_{\leqslant i}[j]\}$;  $\mu_{i+1} = S_{i+1}/n$.
Return $C = (c_1, \ldots, c_m)$.

In Step 2, we *non-deterministically* try all possible choices for $c_i$; the list $\mathcal{L}$ is obtained by taking all possible $C$s that might be returned by this algorithm. Also, $c_i$ is a codeword in $\mathcal{C}^{-S_i}$ but we can think of it as a codeword in $\mathcal{C}$ by the 1-1 correspondence. Different choices for $c_i$ lead to different sets $S_{i+1}$, and hence to different lists $\mathcal{L}_{i+1}$. So the execution of ERASE-DECODE is best viewed as a tree, we formalize this below.

For a received word $R$, $\text{Tree}(R)$ is a tree with $m + 1$ levels. The root is at level 0. A node $v$ at level $i$ is labeled by $C(v) = (c_1, \ldots, c_i)$. It is associated with a set $S(v) \subseteq [n]$ of erasures accumulated so far which has size $\mu(v)n$. The resulting code $\mathcal{C}^{-S(v)}$ has minimum distance $\delta(v)n \geqslant (\delta - \mu(v))n$. We find all codewords in $\mathcal{C}^{-S(v)}$ that are within distance $(\eta - \mu(v))n$ of the received word $r_{i+1}^{-S(v)}$, call this list $\mathcal{L}(v)$. By Lemma 3.2, $\mathcal{L}(v)$ contains at most $\ell(\eta)$ codewords. Each edge leaving $v$ is labelled by a distinct codeword $c_{i+1}$ from $\mathcal{L}(v)$; it is assigned a weight $w(c_{i+1}) = d(c_{i+1}^{-S(v)}, r_{i+1}^{-S(v)})/n$. The weight $w(c) \in [0, 1]$ of an edge indicates how many new erasures that edge contributes. Thus $\mu(v) = w(c_1) + \cdots + w(c_i)$. The leaves at level $m$ correspond to codewords in the list $\mathcal{L}$. There might be no out-edges from $v$ if the list $\mathcal{L}(v)$ is empty. This could result in a leaf node at a level $i < m$ which does not correspond to codewords. Thus the number of leaves in $\text{Tree}(R)$ is an upper bound on the list-size for $R$.

In order to bound the number of leaves, we assign colors to the various edges based on their weights. Let $c$ be an edge leaving the vertex $v$. We color it WHITE if $w(c) < \delta - \eta$, BLUE if $w(c) \geqslant \delta - \eta$ but $w(c) < \frac{\delta(v)}{2}$ and RED if $w(c) \geqslant \frac{\delta(v)}{2}$. WHITE edges correspond to codewords that are very close to the received word, BLUE edges to codewords that are within the unique-decoding radius, and RED edges to codewords beyond the unique decoding radius.

We begin by observing that WHITE edges can only occur if the list is of size 1.

**Lemma 3.3.** *If a vertex $v$ has a WHITE out-edge, then it has no other out-edges.*

*Proof.* Assume that the edge labelled with $c \in \mathcal{L}(v)$ is colored WHITE, so that $d(c, r_{i+1}^{-S(v)}) < (\delta - \eta)n$. Let $c'$ be another codeword in $\mathcal{L}(v)$, so that $d(c', r_{i+1}^{-S(v)}) \leqslant (\eta - \mu(v))n$. Then by the triangle inequality,

$$d(c, c') < (\delta - \eta)n + (\eta - \mu(v))n = (\delta - \mu(v))n \leqslant \delta(v)n$$

But this is a contradiction since $d(c, c') \geqslant \delta(v)n$.                                          $\square$

We observe that BLUE edges do not cause much branching and cannot result in very deep paths.

**Lemma 3.4.** *A vertex can have at most one BLUE edge leaving it. A path from the root to a leaf can have no more than $\lceil \frac{\eta}{\delta - \eta} \rceil$ BLUE edges.*

*Proof.* The first part holds as there can be at most one codeword within the unique decoding radius. Each BLUE edges results in at least $(\delta - \eta)n$ erasures. Therefore, after $\lceil \frac{\eta}{\delta - \eta} \rceil$ BLUE edges, all $\eta n$ errors have been identified, so all remaining edges have to be WHITE.                $\square$

Lastly, we show that RED edges do not give deep paths either, although a vertex can have up to $\ell(\eta)$ RED edges leaving it.

**Lemma 3.5.** *A path from the root to a leaf can have no more than $\lceil \log(\frac{\delta}{\delta - \eta}) \rceil$ RED edges.*

*Proof.* We claim that every RED edge leaving vertex $v$ has weight at least $(\delta - \mu(v))/2$. Indeed, since $c$ is beyond the unique-decoding radius of $\mathcal{C}^{-S(v)}$, $w(c) \geqslant \frac{\delta(v)}{2}$, and the relative distance $\delta(v)$ of the code $\mathcal{C}^{-S(v)}$ at node $v$ satisfies $\delta(v) \geqslant (\delta - \mu(v))n$.

Assume now for contradiction that some path from the root to a leaf contains $k$ red edges for $k > \lceil \log(\frac{\delta}{\delta - \eta}) \rceil$. Suppose that the edges have weights $\rho_1, \dots, \rho_k$ respectively. Contract the BLUE and WHITE edges between successive RED edges into a single edge, whose weight is the sum of weights of the contracted edges. We also do this for the edges before the first RED edge and those after the last RED edge. This gives a path contains $2k + 1$ edges, where the even edges are RED, and the weight of the edges along the path are $\beta_1, \rho_1, \beta_2, \dots, \rho_k, \beta_{k+1}$ respectively. Let $v_i$ be the parent vertex of the $i^{th}$ RED edge for $i \in [k]$. Then we have $\mu(v_1) = \beta_1$ and $\mu(v_i) = \beta_i + \rho_{i-1} + \mu(v_{i-1})$ for $j \geqslant 1$. But since $\rho_{i-1} \geqslant (\delta - \mu(v_{i-1}))/2$ and $\beta_i \geqslant 0$, we get

$$\mu(v_i) \geqslant \frac{\delta + \mu(v_{i-1})}{2}$$

Now a simple induction on $i$ proves that $\mu(v_i) \geqslant \delta(1 - 2^{1-i})$. If we take $i = \lceil \log(\frac{\delta}{\delta - \eta}) \rceil + 1$, then

$$\mu(v_i) \geqslant \delta \left( 1 - \frac{\delta - \eta}{\delta} \right) = \eta.$$

So when we decode at vertex $v_i$, all the error locations have been identified and erased. Hence we are now decoding from $\eta < \delta$ erasures and no errors, so the decoding is unique and error-free. So vertex $v_i$ will have a single WHITE edge leaving it and no RED edges, which is a contradiction. $\qquad \square$

**Theorem 3.6.** *Assume $\eta < \delta$ and let $b = \lceil \frac{\eta}{\delta - \eta} \rceil$, $r = \lceil \log \frac{\delta}{\delta - \eta} \rceil$. Then $\mathrm{Tree}(R)$ has at most $\binom{b+r}{r} \ell(\eta)^r$ leaves (and hence $\ell^{\odot m}(\eta) \leqslant \binom{b+r}{r} \ell(\eta)^r$).*

*Proof.* We first contract the WHITE edges, since they are the only out-edges leaving their parent nodes. This gives a tree with only RED and BLUE edges. Let $t(b, r)$ denote the maximum number of leaves in a tree where each path has at most $b$ BLUE and $r$ RED edges, and each node have have at most one BLUE edge and $\ell(\eta)$ RED edges leaving it. So we have the recursion

$$t(b, r) \leqslant t(b-1, r) + \ell(\eta)t(b, r-1)$$

with the base case $t(b, 0) = 1$. It is easy to check that $t(b, r) \leqslant \binom{b+r}{r} \ell(\eta)^r$. $\qquad \square$

We conclude this section by proving that the condition $\eta < \delta$ is necessary.

**Lemma 3.7.** *If code $\mathcal{C}$ has relative distance exactly $\delta$, then $\ell^{\odot m}(\delta) \geqslant 2^m$.*

*Proof.* Take $c_1, c_2 \in \mathcal{C}$ where $\Delta_q(c_1, c_2) = \delta$. Now take the received word in $\mathcal{C}^{\odot m}$ to be $R = (c_1, \dots, c_1)$. For every $T \subseteq [m]$, let $C_T$ to be the codeword where the $i^{th}$ column equals $c_1$ if $i \in T$ and $c_2$ otherwise. It is easy to show that every such codeword is at distance at most $\delta$ from $R$, showing that $\ell^{\odot m}(\delta) \geqslant 2^m$. $\qquad \square$

3.1. **A Efficient Decoding Algorithm for $\mathcal{C}^{\odot m}$.** Given a received word $R = (r_1, \ldots, r_m)$, for $i \leqslant m$ let $R_{\leqslant i} = (r_1, \ldots, r_i)$.

---

**Algorithm 2.** DecodeC$^{\odot m}$
**Input:** $R \in [q^m]^n, \eta$.
**Output:** List of all $C \in \mathcal{C}^{\odot m}$ so that $\Delta_{q^m}(R, C) \leqslant \eta$.

1. For $i = 1, \ldots, m$
Set $\mathcal{L}_i = $ DecodeC$(r_i, \eta)$.
2. Set $\mathcal{L}_{\leqslant 1} = \mathcal{L}_1$.
3. For $i = 2, \ldots, m$
For $C \in \mathcal{L}_{\leqslant i-1} \times \mathcal{L}_i$,
Add $C$ to $\mathcal{L}_{\leqslant i}$ if $\Delta_{q^m}(C, R_{\leqslant i}) \leqslant \eta$.
4. Return $\mathcal{L}_{\leqslant m}$.

---

**Claim 3.8.** *Assume that* DecodeC$(r, \eta)$ *runs in time $T$. Then* DecodeC$^{\odot m}(R, \eta)$ *returns a list of codewords within distance $\eta$ of $R$ in time $O(mT + m^2 n \ell(\eta) \ell^{\odot m}(\eta))$.*

*Proof.* For any $C$ such that $\Delta(C, R) \leqslant \eta$, it must hold that for every $i$, $\Delta_{q^i}(C_{\leqslant i}, R_{\leqslant i}) \leqslant \eta$. An easy induction on $i$ shows that $C_{\leqslant i} \in \mathcal{L}_{\leqslant i}$, which proves the correctness of the algorithm.

It is clear that Step 1 takes time $O(mT)$. To bound Step 3, we use the following simple observation for all $i \leqslant m$:

$$|\mathcal{L}_{\leqslant i}| \leqslant \ell^{\odot i}(\eta) \leqslant \ell^{\odot m}(\eta).$$

Thus each iteration of the loop in Step 3 requires computing the distance between $R$ and at most $\ell(\eta) \ell^{\odot m}(\eta)$ candidates for $C$.  $\square$

## 4. List Decoding Tensor Products

In this section and the next, $\mathcal{C}_1 \subset \mathbb{F}_q^{n_1}$ and $\mathcal{C}_2 \subset \mathbb{F}_q^{n_2}$ will be linear codes over $\mathbb{F}_q$. Given a matrix $A \in [q]^{n_2 \times n_1}$, and two subsets $S \subseteq [n_2]$ and $T \subseteq [n_1]$, we shall use $A[S, T]$ to denote the submatrix indexed by rows in $S$ and columns in $T$. Further, we will write $A[S, *]$ instead of $A[S, [n_1]]$. Thus the symbol "$*$" when used as an index, denotes the sets $[n_1]$ or $[n_2]$.

Fix a received word $R \in [q]^{n_2 \times n_1}$ and a codeword $C \in \mathcal{C}_2 \otimes \mathcal{C}_1$ so that $\delta(R, C) \leqslant \eta^* - 3\varepsilon$. The advice/guess $A[S, T]$ to the algorithm TensorDecode consists of the values of $C$ on a random submatrix $S \times T$. Given the advice $A$, the TensorDecode algorithm works in four phases, described informally below. This is followed by a formal description of the algorithm in Figure 1. The reader might find Figure 2 helpful to develop intuition about the operation of the various phases.

**Phase 1:** We run the list decoder on each row $s \in S$ to get a new advice string $B[S, \star]$. If there is a codeword in the list that agrees with the advice $A[s, T]$, we set $B[s, \star]$ to be that codeword (ties are broken arbitrarily; or we could declare an erasure if there isn't a unique choice). If there is no such codeword, we set every symbol in $B[s, \star]$ to $\perp$ (which denotes an erasure). Note that the list for row $s$ might contain some codeword other than $C[s, \star]$ that agrees with $C[s, \star]$ on the positions in $T$. In such a case $B[s, \star]$ could be incorrect. As a result, some rows in $B[S, \star]$ agree with the codeword $C$, some of them could be wrong, and the rest are erasures. Claim 5.3 show that with high probability, $(1 - \delta_2 + \varepsilon)|S|$ of the rows are correct, and no more than $\varepsilon|S|$ are incorrect.

**Phase 2:** Viewed columnwise, $B[S, \star]$ gives us advice strings for the co-ordinates $S$ of every column codeword. However, the advice is noisy: it is correct on $(1 - \delta_2 + \varepsilon)$ fraction of co-ordinates within $S$, wrong on an $\varepsilon$ fraction, and $\bot$ on the rest. But since any two codewords in the column code $C_2$ are distance $\delta_2$ apart, in expectation the advice string has more agreement with the correct codeword than any other; thus it is likely to identify the correct codeword from a small list of candidates. We create a new advice string $D[\star, \star]$ by list decoding every column $t \in [n_1]$, and selecting from the list a codeword that disagrees with $B[S, t]$ in less than $\varepsilon$ fraction of co-ordinates. If no such codeword exists we set the column to $\bot$. Claim 5.4 shows that at least $(1 - \delta_1 + 2\varepsilon)$ fraction of columns are correctly decoded to the corresponding columns of $C$ and no more than $\varepsilon$ fraction are incorrectly decoded.

**Phase 3:** Viewed row-wise, $D$ gives an advice string for every row that is correct on at least $(1 - \delta_1 + \varepsilon)n_1$ co-ordinates, wrong on at most $\varepsilon n_1$ and blank on the rest. The advice though noisy is sound: since the code $C_1$ has distance $\delta_1 n_1$, a simple application of the triangle inequality shows that there is a unique codeword which disagrees with $D[s, \star]$ on fewer than $\varepsilon n_1$ co-ordinates, and that is the row codeword $C[S, \star]$. Thus the advice uniquely identifies the correct row codeword. We create a new received word $E$ by list decoding each row and using $D$ to identify the correct codeword in the list, and setting the row to $\bot$ if no such codeword exists. Claim 5.5 shows this step will find the correct codeword on $(1 - \delta_2 + 3\varepsilon)$ fraction of the rows.

**Phase 4:** When viewed column-wise, $E$ gives the correct value of $C$ on $1 - \delta_2 + 3\varepsilon$ fraction of co-ordinates, and is blank on the rest. Crucially, it does not have any incorrect symbols. So now we can recover $C$ by decoding each column from erasures (note that one can uniquely decode $C_2$ from less than a fraction $\delta_2$ of erasures).

It is easy to show that the list decoding radius reached by TensorDecode is the correct one.

**Lemma 4.1.** *For a linear code $C$, $\ell(C^{\otimes 2}, \delta\eta) \geqslant \ell(\eta)$.*

*Proof.* Let $r \in [q]^n$ be a received word with codewords $c_1, \ldots, c_\ell \in C$ within radius $\eta$. Take $c_0$ to be a codeword of minimum weight $\delta$. Define the received word $r' = c_0 \otimes r$. It is easy to see that the codewords $c_i' = c_0 \otimes c_i$ for $i \in [\ell]$ are all within distance $\delta\eta$ from $r'$, which proves the claim. $\qquad \square$

Thus, if list decoding $C$ beyond radius $\eta$ is combinatorially intractable, then so is decoding $C^{\otimes 2}$ beyond radius $\delta\eta$.

---

**Algorithm 3.** TENSORDECODE

**Setup :** Let $\mathsf{Decode}_1$, $\mathsf{Decode}_2$ denote list decoding algorithms for $\mathcal{C}_1$ and $\mathcal{C}_2$, up to error rates $\eta_1$ and $\eta_2$ respectively. Let $\ell_1(\eta_1)$ and $\ell_2(\eta_2)$ be the upper bounds on list size output by $\mathsf{Decode}_1$ and $\mathsf{Decode}_2$ respectively. Fix $\eta^* = \min(\delta_1\eta_2, \delta_2\eta_1)$. Let $\varepsilon > 0$ be a parameter to the algorithm.

**Input :** A received word $R$ such that $\delta(R, \mathcal{C}_2 \otimes \mathcal{C}_1) \leqslant \eta^* - 3\varepsilon$.

**Output :** A list $L$ of all codewords $C \in \mathcal{C}_2 \otimes \mathcal{C}_1$ with $\delta(C, R) \leqslant \eta^* - 3\varepsilon$.

- Pick subsets $S \subseteq [n_2]$ and $T \subseteq [n_1]$ uniformly at random among all subsets of size $m_1$ and $m_2$ respectively.

- For each assignment $A : S \times T \to [q]$

    -- **Phase 1** (Computing $B : S \times [n_1] \to [q]$)
    For each $s \in S$,
        * List decode the row $R[s, *]$:  $\mathcal{L}_s = \mathsf{Decode}_1(R[s, *])$.

        * Set $B[s, *]$ to be an arbitrary codeword $c$ in the list $\mathcal{L}_s$ satisfying $c[T] = A[s, T]$ and $B[s, *] =\perp$ if no such codeword exists.
    Define $S_{fail} = \{s \in S | B[s, *] =\perp\}$ and $S_{success} = S - S_{fail}$.

    -- **Phase 2** (Computing $D : [n_2] \times [n_1] \to [q]$)
    For each $t \in [n_1]$,
        * List decode the column $R[*, t]$:  $\mathcal{L}_t = \mathsf{Decode}_2(R[*, t])$

        * Set $D[*, t]$ to be any codeword $c$ in the list $\mathcal{L}_t$ satisfying $\Delta(c[S_{success}], B[S_{success}, t]) < \varepsilon|S|$. Fix $D[*, t] =\perp$ if no such codeword exists.
    Define $T_{fail} = \{t \in [n_1] | D[*, t] =\perp\}$ and $T_{success} = [n_1] - T_{fail}$.

    -- **Phase 3** (Computing $E : [n_2] \times [n_1] \to [q]$)
    For each $s \in \{n_2\}$,
        * List decode the row $R[s, *]$:  $\mathcal{L}_s = \mathsf{Decode}_1(R[s, *])$.

        * Set $E[s, *]$ to be any codeword $c$ in the list $\mathcal{L}_s$ satisfying $\Delta(c[T_{success}], D[s, T_{success}]) < \varepsilon n_1$. Set $E[s, *] =\perp$ if no such codeword exists.
    Define $U_{fail} = \{s \in [n_2] | E[*, t] =\perp\}$ and $U_{success} = [n_2] - U_{fail}$.

    -- **Phase 4** (Computing $C : [n_2] \times [n_1] \to [q]$)
    For each $t \in \{n_1\}$, Unique decode the column $E[*, t]$ under erasures: $C[*, t] = \mathsf{UniqueDecodeErasures}_2(E[*, t])$

    -- Output $C$ if $\delta(C, R) \leqslant \eta^* - 3\varepsilon$.

---

FIGURE 1. List decoding algorithm for tensor product

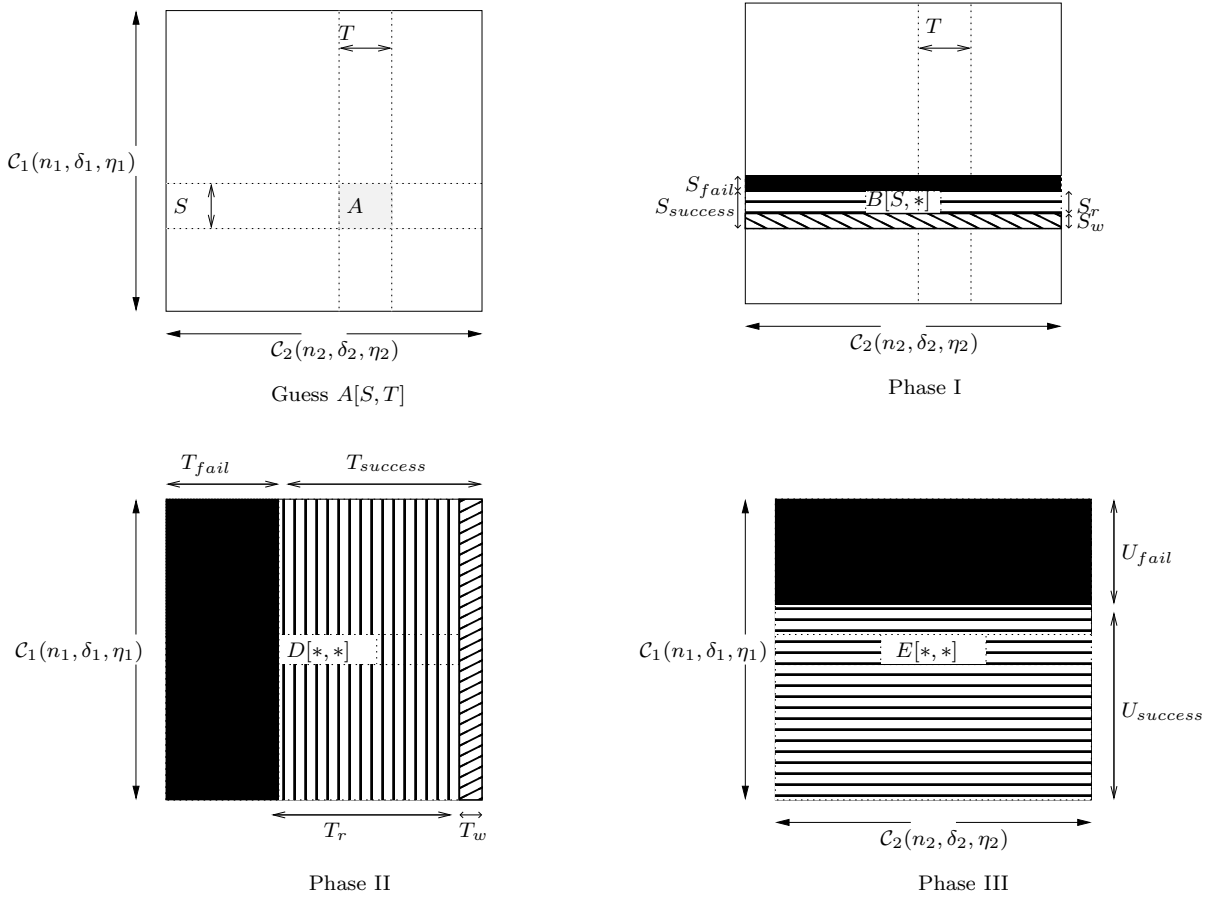

FIGURE 2. Phases of the TENSORDECODE Algorithm

## 5. Analysis of the Tensor Product Decoder

In this section, we will prove the correctness, analyze the list size output and compute running time of the TENSORDECODE algorithm. Towards this goal, we first present a concentration bound concerning sums of samples chosen without replacement.

**Lemma 5.1.** *Let $z_1, z_2, \ldots, z_n$ be real numbers bounded in $[0, 1]$. Let $S \subseteq [n]$ be a uniformly random subset of size $m$. Then,*

$$\Pr\left[\left|\frac{1}{|S|}\sum_{s \in S} z_s - \frac{1}{n}\sum_{i \in [n]} z_i\right| \geqslant \gamma\right] \leqslant p(\gamma, |S|) = 2e^{-2\gamma^2 m}$$

The above concentration bound essentially a restatement of Corollary 1.1 in Serfling's work [27] on sums in sampling without replacement. Henceforth, for the sake of succinctness, we shall use the notation $p(\gamma, m)$ to denote the upper bound $(2e^{-2\gamma^2 m})$ in the above lemma.

Firstly, we will show that for every codeword $C$ close enough to the received word, the algorithm TENSORDECODE returns $C$ with probability $1/4$, given the right advice string $A = C[S, T]$.

**Theorem 5.2.** *For a codeword $C \in \mathcal{C}_2 \otimes \mathcal{C}_1$ within distance $\eta^* - 3\varepsilon$ of the received word $R$, the algorithm* TENSORDECODE *with input $R$, and guess $A = C[S, T]$ returns $C$ with probability at least $1 - p(\varepsilon, m_2) - \ell_1(\eta_1)p(\delta_1, m_1)/\varepsilon - \ell_2(\eta_2)p(\varepsilon, m_2)/\varepsilon$.*

*Proof.* The argument is broken up into four parts (Claims 5.3, 5.4, 5.5, 5.6) each concerning a phase of the TENSORDECODE algorithm. □

5.1. **Phase 1.** In this phase, the algorithm constructed the advice string $B[S, *]$ starting with the guess $A[S, T]$ and list decoding each row in $S$. Of the set of rows $S_{success}$ on which the decoding succeeded, some of them are decoded correctly to the corresponding row in $C$, while some are incorrect. Define sets $S_r, S_w \subseteq S_{success}$ to be the sets of rows that are decoded correctly and incorrectly respectively. Formally,

$$S_r = \{s \in S | B[s, *] = C[s, *]\} \qquad S_w = S_{success} - S_r$$

We make the following claim regarding the fraction of rows decoded correctly.

**Claim 5.3.** *With probability at least $1 - p(\varepsilon, m_2) - \ell_1(\eta_1)p(\delta_1, m_1)/\varepsilon$ over the choice of the sets $S$, $T$*

$$|S_{success}| \geqslant (1 - \delta_2 + 2\varepsilon)|S| \quad |S_r| \geqslant (1 - \delta_2 + \varepsilon)|S| \quad |S_w| \leqslant \varepsilon|S| .$$

*Proof.* Let $S_1$ denote the set of rows in $S$ with fewer than average number of errors. Specifically, $S_1$ is defined as

$$S_1 = \{s \in S | \delta(C[s, *], R[s, *]) \leqslant \eta_1\} .$$

Observe that for each $s \in S_1$, the codeword $C[s, *]$ will be part of the list $\mathcal{L}_s$, obtained by decoding the row $s$. Consequently, for each $s \in S_1$, $B[s, *] \neq \perp$, i.e., $S_1 \subseteq S_{success}$. We shall lower bound the size of $S_{success}$ by the size of $S_1$. Apply Lemma 5.1 with $\{z_i = \delta(C[i, *], R[i, *])\}$ and the set $S$. Since $\delta(C, R) \leqslant \delta_2\eta_1 - 3\varepsilon$, the average of the $z_i$ is less than or equal to $\delta_2\eta_1 - 3\varepsilon$. Thus we get

$$\Pr_S \left[ \delta(C[S, *], R[S, *]) \geqslant \delta_2\eta_1 - 2\varepsilon \right] \leqslant p(\varepsilon, m_2) .$$

Let us suppose $\delta(C[S, *], R[S, *]) \leqslant \delta_2\eta_1 - 2\varepsilon$. By an averaging argument, for at most $\delta_2 - 2\varepsilon$ fraction of the rows $\{C[s, *] | s \in S\}$, the distance $\delta(C[s, *], R[s, *]) > \eta_1$, i.e., $|S_1| \geqslant (1 - \delta_2 + 2\varepsilon)|S|$. This immediately implies the lower bound on size of $S_{success}$.

Now we shall upper bound the probability that a row $s \in S$ is decoded to an incorrect codeword $c \neq C[s, *]$. Fix an $s \in S$ and a codeword $c \in \mathcal{L}_s$ other than $C[s, *]$. The codeword $c$ is chosen, if and only if it agrees with the advice $A[S, T] = C[S, T]$ on the columns in $T$, i.e., $C[s, T] = c[T]$. Applying Lemma 5.1 on the set of real numbers $\left\{z_i = 1\{C[s, i] \neq c[i]\}\right\}$ and the choice of the set $T$, we can conclude

$$\Pr_T \left[ C[s, T] = c[T] \right] \leqslant p(\delta_1, m_1) .$$

By a union bound over all codewords $c \in \mathcal{L}_s$, for any row $s \in S$, the probability of decoding an incorrect codeword is upper bounded by

$$\Pr_T \left[ C[s, *] \neq B[s, *] \wedge B[s, *] \neq \perp \right] \leqslant \ell_1(\eta_1)p(\delta_1, m_1) .$$

Hence in expectation, at most $\ell_1(\eta_1)p(\delta_1, m_1)$ fraction of rows in $S$ are decoded incorrectly, i.e $\mathrm{E}[|S_w|] \leqslant \ell_1(\eta_1)p(\delta_1, m_1)|S|$. Applying Markov's inequality, we get

$$\Pr_T\left[|S_w| \geqslant \varepsilon|S|\right] \leqslant \ell_1(\eta_1)p(\delta_1, m_1)/\varepsilon .$$

Suppose $|S_w| \leqslant \varepsilon|S|$, then observe that

$$|S_r| = |S_{success} - S_w| \geqslant (1 - \delta_2 + \varepsilon)|S| .$$

Thus with probability at least $1 - p(\varepsilon, m_2) - \ell_1(\eta_1)p(\delta_1, m_1)/\varepsilon$ both of the assertions of the claim hold. $\qquad\square$

5.2. **Phase 2.** In Phase 2, the algorithm uses the advice $B[S, *]$ generated in the first phase in order to decode the columns of the matrix. Among the columns $T_{success}$ that are decoded successfully, let $T_r, T_w \subseteq T_{success}$ denote the set of rows that are decoded correctly and incorrectly respectively. Formally, define sets $T_r, T_w$ as follows:

$$T_r = \{t \in [n_1] | D[*, t] = C[*, t]\} \qquad T_w = T_{success} - T_r$$

Given that the advice $B[S, *]$ generated in the first phase is correct on sufficiently many rows, the following claim suggests that at least $1 - \delta_1$ fraction of the columns are decoded successfully, of which most columns agree with codeword $C$. In other words, the advice $D[*, *]$ generated by this phase is a *near-sound* advice string with almost all columns having either the correct value or a failure symbol $\perp$.

**Claim 5.4.** *Conditioned on the event that the assertions of Claim 5.3 hold, With probability at least $1 - \ell_2(\eta_2)p(\varepsilon, m_2)/\varepsilon$ over the choice of the sets $S$ , $T$*

$$|T_{success}| \geqslant (1 - \delta_1 + 3\varepsilon)n_1 \quad |T_r| \geqslant (1 - \delta_1 + 2\varepsilon)n_1 \quad |T_w| \leqslant \varepsilon n_1$$

*Proof.* Along the lines of the proof of Claim 5.3, let $T_1$ be the set of columns with fewer than average fraction of errors. Define the set $T_1$ as follows:

$$T_1 = \{t \in [n_1] | \delta(C[*, t], R[*, t]) \leqslant \eta_2\}$$

By an averaging argument, for at most $\delta_1 - 3\varepsilon$ fraction of the columns $\{C[*, t] | t \in [n_1]\}$, the distance $\delta(C[*, t], R[*, t]) > \eta_2$, i.e., $|T_1| \geqslant (1 - \delta_1 + 3\varepsilon)n_1$.

Observe that for each column $t \in T_1$, the codeword $C[*, t]$ belongs to the list $\mathcal{L}_t$. By Claim 5.3, in Phase 1, at most $\varepsilon|S|$ rows in $S$ were decoded incorrectly, i.e, $|S_w| \leqslant \varepsilon|S|$. Hence the codeword $C[*, t]$ satisfies $\Delta(C[S_{success}, t], B[S_{success}, t]) \leqslant \varepsilon|S|$. Consequently, for each $t \in T_1$, $D[*, t] \neq \perp$, i.e. $T_1 \subseteq T_{success}$. Thus a lower bound on the size of $T_{success}$ is given by $|T_1| \geqslant (1 - \delta_1 + 3\varepsilon)n_1$.

Fix a $t \in [n_1]$. Let us suppose $D[*, t]$ is neither equal to $\perp$ or $C[*, t]$. Thus $D[*, t]$ is a codeword in $\mathcal{C}_2$ such that

$$\Delta(D[S_{success}, t], C[S_{success}, t]) \leqslant \varepsilon|S|$$

By assertion of 5.3,

$$|S_{fail}| \leqslant |S| - |S_{success}| \leqslant (\delta_2 - 2\varepsilon)|S| .$$

From the above inequalities, we can conclude

$$\delta(D[S, t], C[S, t]) \leqslant \delta_2 - \varepsilon .$$

Fix a codeword $c \in \mathcal{L}_t$. By the distance property of the code, we have $\delta(c, C[*, t]) \geqslant \delta_2$. Applying Lemma 5.1 on the set of real numbers $\left\{ z_i = \mathbb{1}\left[ c[i] \neq C[i, t] \right] \right\}$ and the set $S$, we get

$$\Pr_S\left[ \delta(C[S, t], c[S]) \leqslant \delta_2 - \varepsilon \right] \leqslant p(\varepsilon, m_2) \ .$$

By a union bound over all codewords $c \in \mathcal{L}_t$, for a column $t \in T$,

$$\Pr_S\left[ C[*, t] \neq D[*, t] \wedge D[*, t] \neq \perp \right] \leqslant \ell_2(\eta_2) p(\varepsilon, m_2) \ .$$

In other words, the expected size of $T_w$ is at most $\ell_2(\eta_2) p(\varepsilon, m_2)|T|$. Applying Markov's inequality, we get:

$$\Pr_S\left[ |T_w| \geqslant \varepsilon n_1 \right] \leqslant \ell_2(\eta_2) p(\varepsilon, m_2)/\varepsilon \ .$$

To finish the argument, observe that

$$|T_r| = |T_{success} - T_w| \geqslant (1 - \delta_1 + 2\varepsilon) n_1 \ .$$

Thus with probability $1 - \ell_2(\eta_2) p(\varepsilon, m_2)/\varepsilon$ both of the assertions of the claim hold. $\qquad \square$

### 5.3. Phase 3.
This phase converts a *near sound* advice $D[*, *]$ into a *perfectly sound* advice $E[*, *]$ all of whose rows are either the correct codewords or the fail symbol $\perp$. The following claim is a formal statement of this fact.

**Claim 5.5.** *Conditioned on the event that the assertions of Claim 5.4 hold, for each $s \in U_{success}$, $E[s, *] = C[s, *]$ and $|U_{success}| \geqslant (1 - \delta_2 + 3\varepsilon) n_2$.*

*Proof.* For each $s \in [n_2]$, we claim that $C[s, *]$ is the unique codeword satisfying

$$\Delta(c[T_{success}], D[s, T_{success}]) \leqslant \varepsilon n_1.$$

Clearly, this implies that for each row $s \in U_{success}$, we decode the correct codeword, i.e., $E[s, *] = C[s, *]$. For the sake of contradiction, let us suppose there exists $c \neq C[s, *]$ satisfying

$$\Delta(c[T_{success}], D[s, T_{success}]) \leqslant \varepsilon n_1.$$

By triangle inequality, we can conclude

$$\Delta(c[T_{success}], C[s, T_{success}]) \leqslant 2\varepsilon n_1.$$

Further from claim 5.4, $|T_{fail}| \leqslant [n_1] - |T_{success}| \leqslant (\delta_1 - 3\varepsilon) n_1$. This implies

$$\Delta(c, C[s, *]) \leqslant 2\varepsilon n_1 + (\delta_1 - 3\varepsilon) n_1 < \delta_1 n_1.$$

This is a contradiction since $c$ and $C[s, *]$ are two distinct codewords of $\mathcal{C}_1$ that are less than $\delta_1$ apart.

Let $U_1$ denote the set of rows with less than average fraction of errors; formally,

$$U_1 = \left\{ s \in [n_2] \Big| \delta(C[s, *], R[s, *]) \leqslant \eta_1 \right\} \ .$$

By an averaging argument, for at most $\delta_2 - 3\varepsilon$ fraction of the rows of $C$, the distance $\delta(C[s, *], R[s, *]) > \eta_1$, i.e., $|U_1| \geqslant (1 - \delta_2 + 3\varepsilon) n_2$. For each row $s \in U_1$, we have $C[s, *] \in \mathcal{L}_s$. From Claim 5.4, for at most $\varepsilon n_1$ columns in $T_{success}$, $C[*, t] \neq D[*, t]$. Consequently, for each row $s \in U_1$, $E[s, *] \neq \perp$, i.e., $U_1 \subseteq U_{success}$. Hence we get $|U_{success}| \geqslant |U_1| \geqslant (1 - \delta_2 + 3\varepsilon) n_2$. $\quad \square$

5.4. PHASE 4. This is a fairly simple phase where a *perfectly sound* advice $E[*, *]$ is used to completely retrieve the codeword $C$. Specifically, we make the following claim :

**Claim 5.6.** *Conditioned on the event that the assertions of Claim 5.5 hold, the algorithm* TENSORDECODE *outputs the codeword* $C$.

*Proof.* By Claim 5.5, we know $E[s, *] = C[s, *]$ for at least $(1 - \delta_2 + 3\varepsilon)n_2$ rows and $E[s, *] = \perp$ for the remaining rows. Hence for each column $t \in [n_1]$, the UniqueDecodeErasures$_2$ algorithm returns the codeword $C[*, t]$. Thus the algorithm TENSORDECODE returns the codeword $C$. □

**Theorem 5.7.** *Given two codes $\mathcal{C}_1, \mathcal{C}_2$, for every $\varepsilon > 0$, the number of codewords of $\mathcal{C}_2 \otimes \mathcal{C}_1$ within distance $\eta^* = \min(\delta_1\eta_2, \delta_2\eta_1) - 3\varepsilon$ of any received word is bounded by*

$$\ell(\mathcal{C}_2 \otimes \mathcal{C}_1, \eta^*) \leqslant 4q^{\frac{1}{4\delta_1^2\varepsilon^2} \ln \frac{8\ell_1(\eta_1)}{\varepsilon} \ln \frac{8\ell_2(\eta_2)}{\varepsilon}}.$$

*Further, if $\mathcal{C}_1$ and $\mathcal{C}_2$ can be efficiently list decoded up to error rates $\eta_1, \eta_2$ and $\mathcal{C}_2$ is a linear code, then $\mathcal{C}_2 \otimes \mathcal{C}_1$ can be list decoded efficiently up to error rate $\eta^*$. Specifically, if $T$ denotes the time complexity of list decoding $\mathcal{C}_1$ and $\mathcal{C}_2$, then the running time of the list decoding algorithm for $\mathcal{C}_2 \otimes \mathcal{C}_1$ is $O(4q^{\frac{1}{4\delta_1^2\varepsilon^2} \ln \frac{8\ell_1(\eta_1)}{\varepsilon} \ln \frac{8\ell_2(\eta_2)}{\varepsilon}} \times Tn_1n_2)$*

*Proof.* Rewriting the expression for the probability in Theorem 5.2 using Lemma 5.1,

$$1 - p(\varepsilon, m_2) - \frac{\ell_1(\eta_1)p(\delta_1, m_1)}{\varepsilon} - \frac{\ell_2(\eta_2)p(\varepsilon, m_2)}{\varepsilon} = 1 - 2e^{-2\varepsilon^2 m_2} - $$
$$\frac{2}{\varepsilon}\left(\ell_1(\eta_1)e^{-2\delta_1^2 m_1} + \ell_2(\eta_2)e^{-2\varepsilon^2 m_2}\right).$$

Set $m_1 = \frac{1}{2\delta_1^2} \ln \frac{8\ell_1(\eta_1)}{\varepsilon}$ and $m_2 = \frac{1}{2\varepsilon^2} \ln \frac{8\ell_2(\eta_2)}{\varepsilon}$. It is easy to see that the probability of success is at least $\frac{1}{4}$ with this choice of parameters. In other words, with this choice of parameters, any codeword $C$ within distance $\eta^* - 3\varepsilon$ is output with probability at least $\frac{1}{4}$ if the initial guess $A$ is consistent with $C$. Hence, the number of codewords within distance $\eta^* - 3\varepsilon$ from the received word $R$ is

$$\ell(\mathcal{C}_2 \otimes \mathcal{C}_1, \eta^* - 3\varepsilon) \leqslant 4q^{m_1m_2} = 4q^{\frac{1}{4\delta_1^2\varepsilon^2} \ln \frac{8\ell_1(\eta_1)}{\varepsilon} \ln \frac{8\ell_2(\eta_2)}{\varepsilon}}.$$

It is easy to check that the running time of the algorithm is as claimed above. □

**Theorem 5.8.** *Let $\mathcal{C}$ be a linear code with distance $\delta$, list decodable up to an error rate $\eta$. For every $\varepsilon > 0$, the $m$-wise tensor product code $\mathcal{C}^{\otimes m}$ can be list decoded up to an error rate $\delta^{m-1}\eta - \varepsilon$ with a list size $\exp\left(\left(O(\frac{\ln \ell(\eta)/\varepsilon}{\varepsilon^2})\right)^m\right)$.*

*Proof.* Applying Theorem 5.7 with $\mathcal{C}_1 = \mathcal{C}_2 = \mathcal{C}$, we get list size bound at error rate $\delta\eta - 3\varepsilon$ for $\mathcal{C} \otimes \mathcal{C}$. Applying theorem again on $\mathcal{C}^{\otimes 2}$, we get list size bounds at error rate $\delta^2 \times (\delta\eta - 3\varepsilon) - 3\varepsilon = \delta^3\eta - 3\delta^2\varepsilon - 3\varepsilon$. In general, for $\mathcal{C}^{\otimes 2^k}$ let $\eta_{2^k}$ denote the error rate at which we obtain a list size bound. Then,

$$\eta_{2^k} = \delta^{2^k-1}\eta - 3\varepsilon \sum_{i=0}^{k} \delta^{2^i} > \delta^{2^k-1}\eta - \frac{3\varepsilon}{1-\delta^2}$$

For brevity, let us denote by $S_k$ the list size $\ell(\mathcal{C}^{\otimes 2^k}, \eta_{2^k})$. Then from Theorem 5.7, we have the following recursive inequality:

$$\ln S_{k+1} \leqslant \frac{\ln q}{4\delta^2 \varepsilon^2} \ln^2 \frac{4S_k}{\varepsilon} + \ln 4$$

Rewriting the above inequality,

$$\ln \frac{4S_{k+1}}{\varepsilon} \leqslant \frac{\ln q}{4\delta^2 \varepsilon^2} \ln^2 \frac{4S_k}{\varepsilon} + \ln 4 + \ln \frac{4}{\varepsilon} \leqslant \frac{\ln q}{2\delta^2 \varepsilon^2} \ln^2 \frac{4S_k}{\varepsilon}$$

Set $s_k = \ln 4S_k/\varepsilon$ and $a = \frac{\ln q}{2\delta^2 \varepsilon^2}$. Then we have the recurrence relation :

$$s_{k+1} \leqslant a \cdot s_k^2 \qquad s_0 = \ln \frac{4\ell(\eta)}{\varepsilon}$$

Thus we get $s_k \leqslant (\ln 4\ell(\eta)/\varepsilon)^{2^k} a^{2^k - 1} < (a \ln 4\ell(\eta)/\varepsilon)^{2^k}$. Hence we obtain the following list size bound for $m = 2^k$.

$$\ell\left(\mathcal{C}^{\otimes m}, \delta^{m-1}\eta - \frac{3\varepsilon}{1 - \delta^2}\right) \leqslant \exp\left(\left(\frac{\ln q \ln 4\ell(\eta)/\varepsilon}{2\delta^2 \varepsilon^2}\right)^m\right)$$

Rewriting the above expression with $\varepsilon$ in place of $\frac{3\varepsilon}{1-\delta^2}$,

$$\ell(\mathcal{C}^{\otimes m}, \delta^{m-1}\eta - \varepsilon) \leqslant \exp\left(\left(\frac{9\ln q \ln 12\ell(\eta)/\varepsilon(1 - \delta^2)}{2\delta^2(1 - \delta^2)^2 \varepsilon^2}\right)^m\right) .$$

$\square$

## 6. Improved List-size Bounds via Generalized Hamming Weights

In this section, we prove improved list-size bounds on tensor products and interleavings of binary linear codes. This is done by making a connection between the weight-distributions of such codes and the classical coding theoretic notion of Generalized Hamming Weights. This allows us to use the Deletion technique of [12] to reduce the problem of bounding list-sizes to the low-rank case.

We start by introducing the version of the Deletion Lemma that we need. It is a mildly stronger version of the deletion lemma from [12], the graph theoretic view was proposed by Impagliazzo.

**Lemma 6.1** (Deletion lemma). [12] *Let $\mathcal{C} \subset \mathbb{F}_q^n$ be a linear code over $\mathbb{F}_q$. Let $\mathcal{C}' \subseteq \mathcal{C}$ be a (possibly non-linear) subset of codewords so that $c' \in \mathcal{C}'$ iff $-c' \in \mathcal{C}'$, and every codeword $c \in \mathcal{C} \setminus \mathcal{C}'$ has $\mathrm{wt}(c) \geqslant \mu$. Let $\eta = J_q(\mu) - \gamma$ for $\gamma > 0$. Then*

$$\ell(\mathcal{C}, \eta) \leqslant \gamma^{-2}\ell(\mathcal{C}', \eta).$$

*Proof.* Let $r \in \mathbb{F}_q^n$ denote a received word, and let $\mathcal{L} = \{c_1, \ldots, c_t\}$ be the list of all codewords in $\mathcal{C}$ so that $\Delta(r, c_i) \leqslant \eta$. Construct an (undirected) graph $G$ where $V(G) = \{c_1, \ldots, c_t\}$ and $(i, j)$ is an edge if $c_i - c_j \in \mathcal{C}'$. Our goal is to bound $|V(G)|$.

We claim $G$ does not have large independent sets. Let $I = \{c_1, \ldots, c_s\}$ be an independent set. This means that for every $i \neq j \in [s]$, $c_i - c_j \notin \mathcal{C}'$ so $\Delta(c_i, c_j) \geqslant \mu$. But every codeword in $I$ lies within distance $\eta$ of $r$. We now invoke the Johnson bound which states that in a code of distance $\mu$, the list-size at radius $J_q(\mu) - \gamma$ is bounded by $\gamma^{-2}$. This shows that $\alpha(G) \leqslant \gamma^{-2}$.

We claim that the degree of $G$ is bounded by $\ell(\mathcal{C}', \eta)$. Suppose that a vertex $c$ has $d$ neighbors $\{c_1, \ldots, c_d\}$. They can be written as $c + c'_1, \ldots, c + c'_d$ where $c' \in \mathcal{C}'$. Since

$$\Delta(c + c'_i, r) \;=\; \Delta(c'_i, r - c) \;\leqslant\; \eta$$

the codewords $c'_1, \ldots, c'_d$ give us a list of codewords in $\mathcal{C}'$ at distance $\eta$ from the received word $r - c$. Hence $d \leqslant \ell(\mathcal{C}', \eta)$.

Thus $G$ has degree $d(G) \leqslant \ell(\mathcal{C}', \eta)$ and the max independent set size $\alpha(G) \leqslant \gamma^{-2}$. Thus

$$|G| \leqslant \alpha(G) d(G) \leqslant \gamma^{-2} \ell(\mathcal{C}', \eta).$$

$\square$

The Deletion lemma of [12] corresponds to taking $\mathcal{C}'$ to be all codewords for weight less than $\mu$, and using $\ell(\mathcal{C}', \eta) \leqslant |\mathcal{C}'|$. However, in our applications $|\mathcal{C}'|$ will be too large for this to be a useful bound, thus we essentially use the Deletion lemma as a reduction to the low-rank case.

Generalized Hamming Weights (GHWs) arise naturally in the context of list-decoding from erasures [13]. For a vector $v$ of length $n$, $\mathrm{Supp}(v) \subseteq [n]$ denotes the co-ordinates where $v_i \neq 0$. For a vector space $V$, $\mathrm{Supp}(V) = \cup_{v \in V} \mathrm{Supp}(v)$.

**Definition 6.2.** The $r^{th}$ generalized Hamming weight of a linear code $\mathcal{C} \subseteq \mathbb{F}_2^n$ denoted by $\delta_r(\mathcal{C})$ is defined to be $|\mathrm{Supp}(V_k)|/n$ over all $k$-dimensional subspaces $V_k$ of the code $\mathcal{C}$.

Clearly, $\delta_1(\mathcal{C}) = \delta(\mathcal{C})$ is just the minimum distance. The following lower bound on $\delta_r(\mathcal{C})$ which is folklore [15, 33], says that as we consider larger values of $r$, $\delta_r(\mathcal{C})$ approaches $\frac{q}{q-1}\delta$. Thus for binary linear codes, $\delta_r(\mathcal{C})$ approaches $2\delta(\mathcal{C})$ as $r$ grows.

**Lemma 6.3.** *For any linear code $\mathcal{C} \subseteq \mathbb{F}_q^n$ with minimum distance $\delta(\mathcal{C})$ and any $r \geqslant 1$,*

$$\delta_r(\mathcal{C}) \geqslant \frac{q}{q-1} \delta(\mathcal{C}) \left(1 - \frac{1}{q^r}\right).$$

Given a matrix $C \in \mathbb{F}_q^{n \times m}$, let $\mathrm{Rank}(C)$ denote its rank, let $\mathrm{RowSpan}(C)$ be the space spanned by its rows and $\mathrm{ColSpan}(C)$ be the space spanned by its columns. We use the following standard fact from linear algebra:

**Fact 6.4.** *Given $C \in \mathbb{F}_q^{n \times m}$ such that $\mathrm{Rank}(C) = r$, let $\langle v_1, \ldots, v_r \rangle$ be a basis for $\mathrm{RowSpan}(C)$. Then we can write*

$$C = \sum_{s=1}^{r} u_s \otimes v_s$$

*for some vectors $\{u_1, \ldots, u_r\}$ which form a basis for $\mathrm{ColSpan}(C)$.*

6.1. **INTERLEAVED CODES.** In this subsection $\mathcal{C}$ is a binary linear code. We $\ell^{\odot m}(\eta)$ to denote $\ell(\mathcal{C}^{\odot m}, \eta)$. We use $\mathcal{C}_r^{\odot m}$ to denote the sub-code of $\mathcal{C}^{\odot m}$ consisting of codewords of rank at most $r$, and $\ell_r^{\odot m}(\eta)$ for $\ell(\mathcal{C}_r^{\odot m}, \eta)$. The following lemma relates the rank of a codeword to GHWs.

**Lemma 6.5.** *Given $C \in \mathcal{C}^{\odot m}$ such that $\mathrm{Rank}(C) = r$, $\mathrm{wt}(C) \geqslant \delta_r(\mathcal{C})$.*

The lemma holds since $\dim(\mathrm{ColSpan}(C)) = r$ hence its support is at least $\delta_r(\mathcal{C})$. We now apply the deletion argument to reduce the problem of bounding the list-size to the low-rank case.

**Lemma 6.6.** *Let $\mathcal{C}$ be a binary linear code and let $r = \lceil \log \frac{2}{\delta^2} \rceil$. Then for any $\eta \leqslant \delta$, we have*

$$\ell^{\odot m}(\eta) \leqslant \frac{4}{\delta^4} \ell_r^{\odot m}(\eta).$$

*Proof.* It is easy to check that

$$\delta_r \geqslant 2\delta(1 - 2^{-r}) \geqslant 2\delta - \delta^3,$$

$$J(\delta_r) = 1 - \sqrt{1 - \delta_r} > \delta + \frac{\delta^2}{2}.$$

The bound on $J(\delta_r)$ may be verified by observing that

$$J^{-1}(\delta + \frac{\delta^2}{2}) = 2\delta - \delta^3 - \frac{\delta^4}{4} < \delta_r.$$

Let $\mathcal{C}'$ consist of all codewords $C$ where $\mathrm{Rank}(C) \leqslant r$ so that we can take $\mu = \delta_r$. Since $J(\mu) \geqslant \delta + \delta^2/2$ whereas $\eta \leqslant \delta$, we have

$$\gamma = J(\mu) - \eta = (\delta - \eta) + \frac{\delta^2}{2} \geqslant \frac{\delta^2}{2} \ .$$

Applying the Deletion Lemma 6.1, we obtain the desired conclusion $\ell^{\odot m}(\eta) \leqslant \frac{4}{\delta^4} \ell_r^{\odot m}(\eta)$.  □

An immediate corollary of Fact 6.4 is

**Corollary 6.7.** *Given a codeword $C \in \mathcal{C}^{\odot m}$ of rank $r$, let $\{b[1], \ldots, b[r]\}$ be basis for $\mathrm{RowSpan}(C)$. Then $C$ can be written as $C = \sum_{s=1}^{r} c_s \otimes b[s]$ where $c_s \in \mathcal{C}$ for $s \in [r]$.*

Our goal is to reduce the low-rank problem for interleaved codes to the case when $m = r$, by fixing a basis for the row-space. The following lemma narrows the choices for the basis elements to rows that have reasonably large weight.

**Lemma 6.8.** *Let $\varepsilon > 0$ and let $\eta = \delta - \varepsilon$. Let $C \in \mathcal{C}^{\odot m}$ be a rank $r$ codeword and $R$ be a received word such that $\Delta(R, C) \leqslant \eta$. There is a basis $\{b[1], \ldots, b[r]\}$ for $\mathrm{RowSpan}(C)$ where $\mathrm{wt}(C, b[s]) \geqslant \varepsilon 2^{1-r}$ for all $s \in [r]$.*

*Proof.* Let

$$S = \{b \in \mathrm{Span}(C) \mid \mathrm{wt}(R, b) \geqslant \varepsilon 2^{1-r}, b \neq 0^m\}.$$

We claim that $S$ contains a basis for $\mathrm{RowSpan}(C)$, or equivalently $\dim(S) = r$. Assume for contradiction that $\dim(S) = r - 1$, and that $b[1], \ldots, b[r-1]$ is a basis for it. Complete it to a basis for $\mathrm{Span}(C)$ by adding $b[r]$, and let $S' = b[r] + \langle b[1], \ldots, b[r-1] \rangle$. Note that $S'$ is disjoint from $S$.

By Corollary 6.7, we can write $C = \sum_{t=1}^{r} c_t \otimes b[t]$. If $c_r[i] \neq 0$, then $C[i] \in S'$. Since $\mathrm{wt}(c_r) \geqslant \delta n$, it follows that $\delta n$ of the rows of $C$ lie in $S'$; that is $\mathrm{wt}(C, S') \geqslant \delta$. Since $\Delta(C, R) \leqslant \eta$, we have $\mathrm{wt}(R, S') \geqslant \varepsilon$. But since $|S'| = 2^{r-1}$, there must exist $b \in S'$ so that $\mathrm{wt}(R, b) \geqslant \varepsilon 2^{1-r}$. Since $S'$ is disjoint from $S$, this contradicts the definition of the set $S$.  □

**Lemma 6.9.** *Let $\varepsilon > 0$ and let $\eta = \delta - \varepsilon$. Set $r = \lceil \log \frac{2}{\delta^2} \rceil$. We have*

$$\ell^{\odot m}(\eta) \leqslant \frac{2^{r^2+2}}{\delta^4 \varepsilon^r} \ell^{\odot r}(\eta).$$

*Proof.* By Lemma 6.6, it suffices to bound $\ell_r^{\odot m}(\delta - \varepsilon)$ where $r = \lceil \frac{2}{\delta^2} \rceil$. We fix the choice of basis $\{b[1], \ldots, b[r]\}$ for $\mathrm{RowSpan}(C)$. Lemma 6.8 shows that there are at most $2^{r^2} \varepsilon^{-r}$ choices for the basis. We then map $R : \mathbb{F}_2^k \to \mathbb{F}_2^m$ to a received word $R' : \mathbb{F}_2^k \to \{\mathbb{F}_2^r \cup \star\}$ as follows:

$$R'(x) = \begin{cases} (\lambda_1, \ldots, \lambda_r) & \text{if } R(x) = \sum_i \lambda_i b[i] \\ \star & \text{if } R(x) \notin \mathrm{Span}(C) \end{cases}$$

Every $C'$ satisfying $\Delta(R', C') \leqslant \eta$ is in on-to-one correspondence with $C$ so that $\Delta(R, C) \leqslant \eta$ and $\mathrm{Span}(C) = \langle b[1], \ldots, b[r] \rangle$. So the number of such codewords is bounded by $\ell^{\odot r}(\eta)$. $\qquad\square$

Clearly, $\ell^{\odot r}(\eta) \leqslant \ell(\eta)^r$. Plugging this into Lemma 6.9 gives Theorem 2.6. Further improvements on this bound are possible using the analysis of Theorem 3.6 combined with better list-size bounds for decoding binary codes from erasures; we present them is Section 6.3.

The only step which needs $q = 2$ is Lemma 6.6, where we choose $r$ large enough so that $J_2(\delta_r) \geqslant \delta$. This does not have an analogue over $\mathbb{F}_q$ since $\delta_r$ may only increase by a factor $q/(q-1)$, hence there may not be an $r$ such that $J_q(\delta_r) > \delta$. But the step of bounding the number of small-rank codewords works for any field.

6.2. TENSOR PRODUCTS. In this subsection, $\mathcal{C}_1$ and $\mathcal{C}_2$ are binary linear codes. We use $\delta_{i,r}$ to denote the $r^{th}$ generalized weight of $\mathcal{C}_i$. We use $\ell_i(\eta)$ for $\ell(\mathcal{C}_i, \eta)$ and $\ell^{\otimes}(\eta)$ for $\ell(\mathcal{C}_2 \otimes \mathcal{C}_1, \eta)$ and $\ell_r^{\otimes}(\eta)$ for the list-size when we restrict ourselves to codewords of rank at most $r$.

The following lemma relates the weight-distribution of tensor product codes to the generalized Hamming weights of $\mathcal{C}_1$ and $\mathcal{C}_2$. While the lemma is straightforward, we have not found an explicit statement in the literature.

**Lemma 6.10.** *Given $C \in \mathcal{C}_2 \otimes \mathcal{C}_1$ such that $\mathrm{Rank}(C) = r$, $\mathrm{wt}(C) \geqslant 2\delta_1\delta_2(1 - 2^{-r})$.*

*Proof.* The column rank of $C$ is $r$, thus $|\mathrm{Supp}(\mathrm{ColSpan}(C))| \geqslant \delta_{2,r} n_2$. Each of these indices corresponds to a row from $\mathcal{C}_1$ with weight $\delta_1 n_1$, thus overall the codeword has weight at least $\delta_{2,r}\delta_1 n_2 n_1 \geqslant \delta_1\delta_2(1 - 2^{-r})n_1 n_2$. $\qquad\square$

If we let $\mathrm{wt}_r$ denote the minimum weight of a rank $r$ codeword, we have $\mathrm{wt}_r \geqslant 2\delta_1\delta_2(1 - 2^{-r})$. We now show a reduction to the low-rank case for tensor products.

**Lemma 6.11.** *Set $r = \log(\frac{2}{\delta_1\delta_2})$. Then for any $\eta \leqslant \delta_1\delta_2$,*

$$\ell^{\otimes}(\eta) \leqslant \frac{4}{\delta_1^2\delta_2^2} \ell_r^{\otimes}(\eta).$$

*Proof.* We claim that for any $\delta \leqslant \frac{1}{2}$, $J_2(2\delta - \delta^2) > \delta + \delta^2/2$. To prove this, observe that $J_2^{-1}(\delta) = 2\delta(1 - \delta)$. Hence

$$J_2^{-1}(\delta + \delta^2/2) = 2\delta + \delta^2 - 2\delta^2(1 + \delta/2)^2 \ < \ 2\delta - \delta^2.$$

Lemma 6.10 shows that for $r = \log(\frac{2}{\delta_1 \delta_2})$ any codeword $C$ with $\text{Rank}(C) = r$ has weight at least

$$\text{wt}_r = 2\delta_1 \delta_2 (1 - 2^{-r}) \geqslant 2\delta_1 \delta_2 - \delta_1^2 \delta_2^2.$$

We apply the Deletion lemma taking $\mathcal{C}'$ to be all codewords of rank at most $r$, with $\mu = 2\delta_1 \delta_2 - \delta_1^2 \delta_2^2$. Since $J(\mu) > \delta_1 \delta_2 + \delta_1^2 \delta_2^2 / 2$, whereas $\eta \leqslant \delta_1 \delta_2$, we can take $\gamma = \delta_1^2 \delta_2^2 / 2$, which gives the desired bound. $\qquad\square$

A corollary of Fact 6.4 for tensor product codes is:

**Corollary 6.12.** *Let $C \in \mathcal{C}_2 \otimes \mathcal{C}_1$ be a codeword of rank $r$, and let $\langle v_1, \ldots, v_r \rangle = \text{RowSpan}(C)$. Then $C$ can be written as $C = \sum_{s=1}^{r} u_s \otimes v_s$ where $\langle u_1, \ldots, u_r \rangle = \text{ColSpan}(C)$.*

Fix a received word $R$, which we wish to decode from $\eta^\star - \varepsilon$ fraction of error where $\eta^\star = \min(\eta_1 \delta_2, \eta_2 \delta_1)$ and $\varepsilon > 0$. By decoding each row up to radius $\eta_1$, we get lists $\mathcal{L}_1, \ldots, \mathcal{L}_{n_2}$ of codewords from $\mathcal{C}_1$ each of size at most $\ell_1(\eta_1)$. By decoding each column up to radius $\eta_2$, we get lists $\mathcal{L}'_1, \ldots, \mathcal{L}'_{n_1}$ of codewords from $\mathcal{C}_2$ each of size at most $\ell_2(\eta_2)$. The following lemma gives an analogue of Lemma 6.8, restricting the choice of basis vectors to those that occurs relatively frequently among the lists.

**Lemma 6.13.** *Let $C \in \mathcal{C}_2 \otimes \mathcal{C}_1$ be a rank $r$ codeword such that $\Delta(R, C) \leqslant \eta^\star - \varepsilon$. There is a basis $V = \{v_1, \ldots, v_r\}$ for $\text{RowSpan}(C)$ where each $v_i$ occurs in at least $\varepsilon 2^{1-r} n_2$ of the lists $\mathcal{L}_i$.*

*Proof.* Consider the set of codewords $S = \{v \in C_1\}$ which occur in the row lists at least $\varepsilon 2^{1-r} n_2$ times. We claim that $S$ contains a basis for $\text{RowSpan}(C)$. Assume for contradiction that it only spans an $r - 1$ dimensional subspace. Choose $\{v_1, \ldots, v_{r-1}\}$ which form a basis for it and complete it to a basis by adding $v_r$. Define the set $S' = v_r + \langle v_1, \ldots, v_{r-1} \rangle$. Now by Corollary 6.12, we can write $C$ in the form $C = \sum_{s=1}^{r} u_s \otimes v_s$ for some $u_1, \ldots, u_r \in S$ which span $\text{ColSpan}(C)$.

Note that $\text{wt}(u_r) \geqslant \delta_2$, hence at least $\delta_2 n_2$ rows (corresponding to indices in the support of $u_r$) come from the set $S'$, call this set $A \subset [n_2]$. Since the error rate is $\eta^\star - \varepsilon \leqslant \delta_2 \eta_1 - \varepsilon$, it must be the case that for some subset $B \subseteq A$ of rows where $|B| \geqslant \varepsilon n_2$, the error rate on those rows is less than $\eta_1$; else the overall error rate is at least $(\delta_2 - \varepsilon)\eta_1 > \delta_2 \eta_1 - \varepsilon \geqslant \eta^\star - \varepsilon$. List decoding rows in $B$ up to radius $\eta_1$ recovers the corresponding row vector from $C$. So a vector from $S'$ occurs in all lists for rows in $B$. Hence one of these vectors has to occur with frequency at least $\varepsilon 2^{1-r} n_2$, but this contradicts the fact that $S$ and $S'$ are disjoint. $\qquad\square$

Similarly, let $T$ denote the set of vector $u \in C_2$ which occur in in at least $\varepsilon 2^{1-r} n_1$ of the lists $\mathcal{L}'_i$. One can show that $T$ contains a basis $U = \{u_1, \ldots, u_r\}$ for $\text{ColSpan}(C)$.

**Lemma 6.14.** *We have*

$$\ell_r^\otimes(\eta^\star - \varepsilon) \leqslant 2^{4r^2} \varepsilon^{-2r} \ell_1(\eta_1)^r \ell_2(\eta_2)^r.$$

*Proof.* Note that $|S| \leqslant 2^{r-1} \ell_1(\eta_1) \varepsilon^{-1}$, and $|T| \leqslant 2^{r-1} \ell_2(\eta_2) \varepsilon^{-1}$. We choose $r$ basis vectors from $S$ and $T$ respectively as bases for $\text{RowSpan}(C)$ and $\text{ColSpan}(C)$, for which there are at most $|S|^r |T|^r$ choices. We then choose $r$ row vectors $\{v_1, \ldots, v_r\}$ from $\text{RowSpan}(C)$ and $r$ column vectors $\{u_1, \ldots, u_r\}$ from $\text{ColSpan}(C)$ so that $C = \sum_{s=1}^{r} u_i \otimes v_i$. This gives $2^{2r^2}$

additional choices. Thus $\ell_2^{\otimes}(\eta^\star - \varepsilon)$ can be bounded by $2^{2r^2}|S|^r|T|^r$, which gives the desired bound. □

Putting together Lemmas 6.14 and 6.11, we have proved the following theorem.

**Theorem 6.15.** *Let* $r = \lceil \log(\frac{2}{\delta_1 \delta_2}) \rceil$, $\eta_2 \leqslant \delta_1, \eta_2 \leqslant \delta_2$ *and* $\eta^\star = \min(\eta_1 \delta_2, \eta_2 \delta_1)$. *Then there exist constants* $c_1, c_2$ *so that for any* $\varepsilon > 0$.

$$\ell^{\otimes}(\eta^\star - \varepsilon) \leqslant c_1 2^{c_2 r^2} \ell_1(\eta_1)^r \ell_2(\eta_2)^r \varepsilon^{-2r}.$$

6.3. **Further Improvements for Interleaved Codes.** Theorem 2.6 was proved by first reducing to the rank $r$ case for constant $r$, then reducing to $m = r$ by fixing a basis, and using the trivial upper bound $\ell^{\odot r}(\eta) \leqslant \ell(\eta)^r$. One can improve on this last bound using the analysis of Theorem 3.6 combined with better list-size bounds for decoding binary codes from erasures.

**Theorem 6.16.** *For any binary linear code* $\mathcal{C}$ *of relative distance* $\delta$, *let* $r = \lceil \log \frac{2}{\delta^2} \rceil$. *For any* $\eta < \delta$

$$(6.1) \qquad \ell^{\odot m}(\eta) \leqslant \frac{2^{2r^2}}{\delta^4(\delta - \eta)^r} \prod_{k=0}^{r-1} \ell\left(\eta - \frac{\delta(1 - 2^{-k})}{2}\right)$$

For a binary error correcting code with relative distance $\delta$ and $\mu \leqslant \eta < \delta$, we let $\ell'(\mu, \eta - \mu)$ denote the list-size for $\mathcal{C}$ with $\mu$ erasures and $\eta - \mu$ errors.

**Lemma 6.17.** *For any* $\mu \leqslant \eta$, *we have*

$$\ell'(\mu, \eta - \mu) \leqslant 2 \cdot \ell(\eta - \mu/2).$$

*Proof.* Let $r \in \{0, 1, \star\}^n$ be a received word where $\mathrm{wt}(r, \star) \geqslant \mu$. Let $\mathcal{L} = \{c_1, \cdots, c_k\}$ denote all codewords of $\mathcal{C}$ that satisfy $\Delta(c_i, r) \leqslant \eta$. Set the erased positions of $r$ at random in $\{0, 1\}$, call this received word $r'$. Then $\Pr[\Delta(r', c_i) \leqslant \eta - \mu/2] \geqslant \frac{1}{2}$. Thus, in expectation over the settings of the erased bits, $k/2$ of the codewords from $\mathcal{L}$ satisfy $\Delta(r', c_i) \leqslant \eta - \mu/2$. Fixing one such choice of $r$:

$$\frac{\ell'(\mu, \eta - \mu)}{2} \leqslant \ell(\eta - \mu/2).$$

□

The following lemma completes the proof of Theorem 6.16.

**Lemma 6.18.** *Let* $\mathcal{C}$ *be a binary code with distance* $\delta$. *Then for any* $\eta < \delta$ *we have*

$$(6.2) \qquad \ell^{\odot r}(\eta) \leqslant r 2^r \prod_{k=0}^{r-1} \ell\left(\eta - \frac{\delta(1 - 2^{-k})}{2}\right).$$

*Proof.* We run Algorithm 1 on the received word $R$ and build a tree. We mark each edge leaving $v$ BLUE if it lies within the unique-decoding radius (for the code being decoded at $v$) and RED otherwise. A simple induction shows that after $k$ RED edges, we have at least $\delta(1 - 2^{-k})$ erasures. Thus, after $k$ RED edges, the degree of the tree drops to

$$\ell'(\eta, \delta(1 - 2^{-k})) \leqslant 2\ell\left(\eta - \frac{\delta(1 - 2^{-k})}{2}\right).$$

Solving the recursion for the number of leaves shows that

$$\ell^{\odot r}(\eta) \leqslant r2^r \prod_{k=0}^{r-1} \ell\left(\eta - \frac{\delta(1 - 2^{-k})}{2}\right).$$

$\square$

Equation 6.2 allows us to replace $\ell(\eta)^r$ term in Equation 2.3 with the product $\ell(\eta)\ell(\eta - \delta/4)\ell(\eta - 3\delta/8)\cdots$. This advantage is pronounced if the list size decreases rapidly as the radius shrinks; which happens if we use the Johnson bound to bound the list-size. We finish the proof of Theorem 2.7.

*Proof of Theorem 2.7.*
Set $\eta = J_2(\delta) - \varepsilon$. Note that $\delta - \eta \geqslant \delta - J_2(\delta)$. Let $r = \lceil \log \frac{2}{\delta^2} \rceil$. Using Lemma 6.9, we get

$$(6.3) \qquad \ell^{\odot m}(J_2(\delta) - \varepsilon) \leqslant \frac{4}{\delta^4} \frac{2^{r^2}}{(\delta - J_2(\delta))^r} \ell^{\odot r}(J_2(\delta) - \varepsilon) = c_\delta \ell^{\odot r}(J_2(\delta) - \varepsilon) ,$$

for some constant $c_\delta$ that only depends on $\delta < 1/2$. Applying Lemma 6.18, we conclude that

$$\ell^{\odot r}(J_2(\delta) - \varepsilon) \leqslant r2^r \prod_{k=0}^{r-1} \ell\left(J_2(\delta) - \gamma_k\right) ,$$

where $\gamma_k = \frac{\delta(1 - 2^{-k})}{2} + \varepsilon$. The Johnson bound states that $\ell(J_2(\delta) - \gamma) \leqslant \gamma^{-2}$. For $k \geqslant 1$, we can lower bound $\gamma_k$ by $\frac{\delta(1 - 2^{-k})}{2}$. Hence we have

$$(6.4) \qquad \ell^{\odot r}(J_2(\delta) - \varepsilon) \leqslant r2^r \varepsilon^{-2} \prod_{k=1}^{r-1} \frac{4}{\delta^2 (1 - 2^{-k})^2} \leqslant c'_\delta \varepsilon^{-2}.$$

Combining Equations 6.3 and 6.4 gives the desired result. $\square$

# 7. List Decoding Linear Transformations

In the previous sections, we have developed techniques for decoding generic interleaved codes based on generalized Hamming weights and decoding from erasures. These can be applied to get sharper bounds for specific families of codes. We illustrate this in the case of linear transformations.

The problem of list decoding linear transformations is equivalent to list decoding interleaved Hadamard codes. Equivalently, one can think of the message as a matrix $M$ over $\mathbb{F}_q$ of dimension $k \times m$, encoded by the values $x^t M$ for every $x \in \mathbb{F}_q^k$. Thus the encoding is a matrix $C$ of dimension $q^k \times m$ where each column is a codeword in the Hadamard code. Recall the well known list size bound $\ell(1 - \frac{1}{q} - \varepsilon) \leqslant O(\varepsilon^{-2})$ for Hadamard codes over $\mathbb{F}_q$.

Consider the space $\mathrm{Lin}(\mathbb{F}_q, k, m)$ of all linear transformations $L : \mathbb{F}_q^k \to \mathbb{F}_q^m$. We use $\ell^{\odot m}(\mathrm{Had}_q, \eta)$ instead of to denote the denote the list-size for $\mathrm{Lin}(\mathbb{F}_q, k, m)$ at distance $\eta$. We let $\mathrm{Lin}_r(\mathbb{F}_q, k, m)$ denote the space of all linear transformations of rank at most $r$ in $\mathrm{Lin}(\mathbb{F}_q, k, m)$ and let $\ell_r^{\odot m}(\mathrm{Had}_q, \eta)$ denote the list-size for $\mathrm{Lin}_r(\mathbb{F}_q, k, m)$.

## 7.1. Linear Transformations over $\mathbb{F}_2$.

We first use the Deletion lemma to reduce the task of proving list-size bounds for $\mathrm{Lin}(\mathbb{F}_2, k, m)$ to that of proving bounds for $\mathrm{Lin}_2(\mathbb{F}_2, k, m)$. In this subsection, we use $\ell^{\odot m}(\eta)$ to denote $\ell^{\odot m}(\mathrm{Had}_2, \eta)$ and $\ell_r^{\odot m}(\eta)$ to denote $\ell_r^{\odot m}(\mathrm{Had}_2, \eta)$.

**Lemma 7.1.** *For any $\eta < 1/2$, we have $\ell^{\odot m}(\eta) \leqslant C \ell_2^{\odot m}(\eta)$ where $C$ is an absolute constant.*

*Proof.* We take $\mathcal{C}'$ to consist of all linear transformations of rank at most 2. Thus for any codeword $L \notin \mathcal{C}'$, $\mathrm{Rank}(L) \geqslant 3$, hence $\mathrm{wt}(L) \geqslant \frac{7}{8}$, so we take $\mu = \frac{7}{8}$. Since $J(7/8) > 0.65$ and $\eta < \frac{1}{2}$, we can apply the Deletion lemma with $\gamma = 0.15$ to conclude that $\ell^{\odot m}(\eta) \leqslant C \ell_2^{\odot m}(\eta)$. $\qquad\square$

To bound $\ell_2^{\odot m}(\eta)$, we start by bounding $\ell_1^{\odot m}(\eta)$, which is the maximum number of rank 1 linear transformations within distance $\eta$ of the received word $R$. We need some facts about the structure of $L(x)$ when $L$ is a linear transformation of rank $r$. Recall that we use $l[1], \ldots, l[k] \in \mathbb{F}_2^m$ for the rows of $L$, thought of as vectors from $\mathbb{F}_2^m$. Let $\mathrm{RowSpan}(L) \subseteq \mathbb{F}_2^m$ denote the $r$-dimensional space spanned by these vectors. We write $\mathrm{RowSpan}(L) = \langle l'[1], \ldots, l'[r] \rangle$ to denote the fact that $l'[1], \ldots, l'[r]$ form a basis for $\mathrm{RowSpan}(L)$. For a function $r : \mathbb{F}_2^k \to \mathbb{F}_2^m$ and a vector $v \in \mathbb{F}_2^m$, we define $\mathrm{wt}(R, v) = \mathrm{Pr}_x[R(x) = v]$.

**Lemma 7.2.** *For $L \in \mathrm{Lin}_r(F_2, k, m)$ and any $v \in \mathrm{RowSpan}(L)$, $\mathrm{wt}(L, v) = 2^{-r}$.*

*Proof.* The $j^{th}$ column of $L$ defines a linear function $l_j(x) = x^t l_j$ from $\mathbb{F}_2^k \to \mathbb{F}_2$. We have $L(x) = (l_1(x), \ldots, l_t(x))$. Let us pick a basis for the columns, assume that this basis is $l_1, \ldots, l_r$. Let $v_{\leqslant r}$ denote the projection of $v$ onto the first $r$ co-ordinates. We have $\mathrm{Pr}_x[(l_1(x), \ldots, l_r(x)) = v_{\leqslant r}] = 2^{-r}$, which implies the claim. $\qquad\square$

### 7.1.1. *Bounding rank 1 linear transformations.*

The above lemma implies, in particular, that if $\mathrm{Rank}(L) = 1$ and $\mathrm{RowSpan}(L) = \langle l'[1] \rangle$, then $\mathrm{wt}(L, l'[1]) = \mathrm{wt}(L, 0^k) = \frac{1}{2}$.

**Corollary 7.3.** *Let $L \in \mathrm{Lin}_1(\mathbb{F}_2, k, m)$ and $R : \mathbb{F}_2^k \to \mathbb{F}_2^m$ be such that $\Delta(R, L) \leqslant \frac{1}{2} - \varepsilon$. If $\mathrm{RowSpan}(L) = \langle l'[1] \rangle$, then $\mathrm{wt}(R, l'[1]) \geqslant \varepsilon$.*

This narrows the choice of basis vectors for $\mathrm{RowSpan}(L)$ to at most $\frac{1}{\varepsilon}$ vectors where $\mathrm{wt}(L, l'[1]) \geqslant \varepsilon$. Once we fix $l'[1]$, the problem reduces to Hadamard decoding (or rather $m = 1$). Given $R : \mathbb{F}_2^k \to \mathbb{F}_2^m$, we define $r : \mathbb{F}_2^k \to \{\mathbb{F}_2 \cup \star\}$ as follows:

$$r(x) = \begin{cases} 0 \text{ if } R(x) = 0^m, \\ 1 \text{ if } R(x) = l'[1], \\ \star \quad \text{otherwise.} \end{cases}$$

Setting $r(x) = \star$ denotes an erasure; since $R(x) \notin \mathrm{RowSpan}(L)$, we know there is an error at index $x$. Given a Hadamard codeword $l : \mathbb{F}_2^k \to \mathbb{F}_2$, if we define $L \in \mathrm{Lin}_1(\mathbb{F}_2, k, m)$ by reversing

the substitution:

$$L(x) = \begin{cases} l'[1] \text{ if } l(x) = 1, \\ 0^m \text{ if } l(x) = 0 \end{cases}$$

then it follows that $\Delta(R, L) = \Delta(r, l)$. This proves the following claim.

**Lemma 7.4.** *The linear transformations* $L \in \mathrm{Lin}_1(F_2, k, m)$ *so that* $\Delta(R, L) < \frac{1}{2} - \varepsilon$ *and* $\mathrm{RowSpan}(L) = \langle l'[1] \rangle$ *are in one-to-one correspondence with Hadamard codewords* $l$ *so that* $\Delta(r, l) < \frac{1}{2} - \varepsilon$.

Since there can be at most $O(\frac{1}{\varepsilon^2})$ codewords of the Hadamard code within distance $\frac{1}{2} - \varepsilon$, and at most $\frac{1}{\varepsilon}$ choices for $l'[1]$, this suffices to prove a bound of $O(\frac{1}{\varepsilon^3})$. We can improve this to $O(\frac{1}{\varepsilon^2})$ by observing that if there are many choices for $l'[1]$, then each of them is likely to result in fewer codewords. This relies on the following Lemma about Hadamard decoding with erasures.

**Lemma 7.5.** *Given* $r : \mathbb{F}_2^k \to \{\mathbb{F}_2 \cup \star\}$ *so that* $\mathrm{wt}(R, \star) \geqslant \eta$, *the number of codewords* $l$ *such that* $\Delta(r, l) < \varepsilon$ *is bounded by* $\frac{2}{(\eta + 2\varepsilon)^2}$.

*Proof.* We use the well known fact that the number of codewords so that $\Delta(r, l) < \frac{1}{2} - \varepsilon$ is bounded by $\frac{1}{4\varepsilon^2}$. So assume there are $s$ codewords $l_1, \ldots, l_s$. Consider setting each erasure to a random value in $\mathbb{F}_2$. For any $l$, with probability $\frac{1}{2}$ we have $\Delta(r, l) \leqslant \frac{1}{2} - \frac{\eta}{2} - \varepsilon$. Thus in expectation, $\frac{s}{2}$ of the codewords will now satisfy $\Delta(\bar{l}, r) < \frac{1}{2} - \frac{\eta}{2} - \varepsilon$. Fix one such setting of the erasures. But now we have

$$\frac{s}{2} \leqslant \frac{1}{(\eta + 2\varepsilon)^2} \implies s \leqslant \frac{2}{(\eta + 2\varepsilon)^2}.$$

□

We can now show an $O(\varepsilon^{-2})$ bound on rank 1 transformations.

**Lemma 7.6.** *For any* $\varepsilon > 0$, *we have*

$$\ell_1^{\odot m}(\frac{1}{2} - \varepsilon) \leqslant \frac{1}{2\varepsilon^2}.$$

*Proof.* Assume that the condition $\mathrm{wt}(R, l') > \varepsilon$ holds for $t$ non-zero row vectors $l' \in \mathbb{F}_2^m$. Fix one such choice of $l'$. We then erase $\eta \geqslant (t - 1)\varepsilon$ positions, so by Lemma 7.5 the number of candidates is no more than $\frac{2}{(t+1)^2\varepsilon^2}$. Thus overall the list-size is bounded by $\frac{2t}{(t+1)^2\varepsilon^2} \leqslant \frac{1}{2\varepsilon^2}$. □

7.1.2. *Bounding rank* 2 *linear transformations.* We now proceed to analyzing the list-size for rank 2 linear transformations. We begin with the analog of Corollary 7.3 for the rank two case.

**Corollary 7.7.** *Let* $L \in \mathrm{Lin}_2(\mathbb{F}_2, k, m)$ *and* $R : \mathbb{F}_2^k \to \mathbb{F}_2^m$ *be such that* $\Delta(R, L) \leqslant \frac{1}{2} - \varepsilon$. *There is a basis* $\{u, v\}$ *for* $\mathrm{RowSpan}(L)$ *such that* $\mathrm{wt}(R, u)$ *and* $\mathrm{wt}(R, v)$ *are both at least* $\frac{\varepsilon}{2}$.

*Proof.* Note that $\mathrm{wt}(L, l[1]), \mathrm{wt}(L, l[2])$ and $\mathrm{wt}(L, l[1] + l[2])$ are all exactly $\frac{1}{4}$. If two of the symbols $\{l[1], l[2], l[1]+l[2]\}$ have weight less than $\frac{\varepsilon}{2}$ in $R$, then $\Delta(R, L) > 2(\frac{1}{4} - \frac{\varepsilon}{2}) = \frac{1}{2} - \varepsilon$. □

In fact, we can assume that $\mathrm{wt}(R, u) \geqslant \frac{1}{12}$ and $\mathrm{wt}(R, v) \geqslant \frac{\varepsilon}{2}$, but we will not use this claim. Once we pick $u$ and $v$ we reduce to the case $m = 2$ by defining $r : \mathbb{F}_2^k \to \{\mathbb{F}_2^2 \cup \star\}$ as follows:

$$r(x) = \begin{cases} (\lambda, \mu) \text{ if } R(x) = \lambda u + \mu v, \\ \star \quad \text{otherwise.} \end{cases}$$

We have the following analogue of Lemma 7.4.

**Lemma 7.8.** *The linear transformations $L \in \mathrm{Lin}_2(\mathbb{F}_2, k, m)$ so that $\Delta(L, R) < \frac{1}{2} - \varepsilon$ are in one-to-one correspondence with linear transformations $l \in \mathrm{Lin}_2(\mathbb{F}_2, k, 2)$ so that $\Delta(l, r) < \frac{1}{2} - \varepsilon$.*

**Lemma 7.9.** *Let $r : \mathbb{F}_2^k \to \{\mathbb{F}_2^2 \cup \star\}$ be such that $\mathrm{wt}(r, \star) \geqslant \eta$. The number of linear transformations $l \in \mathrm{Lin}_2(\mathbb{F}_2, k, 2)$ so that $\Delta(l, r) < \frac{1}{2} - \varepsilon$ is bounded by $\frac{100}{(\eta + 2\varepsilon)^2}$.*

*Proof.* Let $l = (l_1, l_2)$ where each $l_i : \mathbb{F}_2^k \to \mathbb{F}_2$. We proceed as in Theorem 2.5, erasing errors whenever they are located. We run Hadamard decoding on the first column which gives a list $\mathcal{L}_1$ of $\frac{2}{(\eta + 2\varepsilon)^2}$ candidates for $\ell_1$. We then choose $c \in \mathcal{L}_1$, erase the positions where it differs from $R$ and list decode the resulting word to get candidates for $l_2$.

At most one codeword $c \in \mathcal{L}_1$ can have error rate less than $\frac{1}{4}$. If we choose $l_1 = c$, we have no more than $\frac{2}{(\eta + 2\varepsilon)^2}$ choices for the second column. The other codewords in $\mathcal{L}_1$ which are beyond distance $\frac{1}{4}$ result in $\eta \geqslant \frac{1}{4}$ erasures, hence a list-size of $C \leqslant 32$ for the second column. So the overall list size is bounded by

$$1 \cdot \frac{2}{(\eta + 2\varepsilon)^2} + \frac{2}{(\eta + 2\varepsilon)^2} \cdot 32 \leqslant \frac{100}{(\eta + 2\varepsilon)^2}.$$

$\square$

We are now ready to bound the number of rank 2 transformations.

**Lemma 7.10.** *There is a constant $C'$ so that for any $\varepsilon > 0$, we have*

$$\ell_2^{\odot m}(\tfrac{1}{2} - \varepsilon) \leqslant \frac{C'}{\varepsilon^2}.$$

*Proof.* Assume there are $t \geqslant 2$ non-zero vectors $v \in \mathbb{F}_2^m$ so that $\mathrm{wt}(v) \geqslant \frac{\varepsilon}{2}$. By Corollary 7.7, each there is a basis $\{u, v\}$ of $L$ where $\mathrm{wt}(R, u)$ and $\mathrm{wt}(R, v)$ are both at least $\frac{\varepsilon}{2}$. Once we fix this basis, we erase all positions containing vectors other than $u, v$ and $u + v$. This results in $\eta \geqslant (t - 3)\varepsilon$ erasures. Thus by Lemma 7.9, the number of $L$ such that $\mathrm{RowSpan}(L) = \langle u, v \rangle$ is bounded by

$$\frac{100}{(\eta + 2\varepsilon)^2} \leqslant \frac{100}{(t - 1)^2 \varepsilon^2}.$$

Since there are $\binom{t}{2}$ choices for $\{u, v\}$, we get an overall bound of

$$\binom{t}{2} \cdot \frac{100}{(t - 1)^2 \varepsilon^2} = \frac{t}{2(t - 1)} \frac{100}{\varepsilon^2} \leqslant \frac{100}{\varepsilon^2}$$

If there are $t < 2$ nonzero vectors $v \in \mathbb{F}_2^m$ with $\mathrm{wt}(v) \geqslant \frac{\varepsilon}{2}$, then it corresponds to the rank one case. Thus, adding bounds for rank 1 and rank 2 cases, we have $\ell_2^{\odot m}(\frac{1}{2} - \varepsilon) \leqslant \frac{100}{\varepsilon^2} + \frac{1}{2\varepsilon^2} = \frac{C'}{\varepsilon^2}$. $\square$

7.1.3. *Final list-size bound.* Combining Lemma 7.10 with Lemma 7.1, we conclude that for some absolute constant $C > 0$,

$$\ell^{\odot m}(\frac{1}{2} - \varepsilon) \leqslant \frac{C}{\varepsilon^2} \ ,$$

thus completing the proof of Theorem 2.9.

## 7.2. LINEAR TRANSFORMATIONS OVER $\mathbb{F}_q$.

In this subsection, we use $\ell^{\odot m}(\eta)$ to denote $\ell^{\odot m}(\mathrm{Had}_q, \eta)$ and $\ell_r^{\odot m}(\eta)$ to denote $\ell_r^{\odot m}(\mathrm{Had}_q, \eta)$. One can repeat the above proof of Theorem 2.9 from the previous subsection for any $\mathbb{F}_q$, and show that for some absolute constant $c$,

$$\ell^{\odot m}(\mathrm{Had}_q, 1 - 1/q - \varepsilon) \leqslant \frac{cq^6}{\varepsilon^2} \ .$$

This gives an asymptotically tight bound for constant $q$. In this subsection, our goal will be to remove the dependence on $q$. We will do so at the expense of a worse dependence (of $\varepsilon^{-5}$) of the bound on $\varepsilon$.

We begin with a deletion argument analogous to Lemma 7.1. The difference is that we only delete rank 1 transformations, and the multiplicative overhead is $O(\varepsilon^{-2})$ as opposed to an absolute constant.

**Lemma 7.11.** *For $\eta = \frac{q-1}{q} - \varepsilon$, we have $\ell^{\odot m}(\eta) \leqslant \frac{C}{\varepsilon^2} \ell_1^{\odot m}(\eta)$.*

*Proof.* We take $\mathcal{C}'$ to consist of all linear transformations of rank at most 1. So we can take $\mu = 1 - \frac{1}{q^2}$ and $J(\mu) = 1 - \frac{1}{q}$. Thus we can apply the Deletion lemma with $\eta = \varepsilon$ to conclude that $\ell^{\odot m}(\eta) \leqslant \frac{1}{\varepsilon^2} \ell_1^{\odot m}(\eta)$. $\qquad\square$

Given $v \in \mathbb{F}_q^m$ we now define $\mathrm{wt}(R, v) = \sum_{\mu \in \mathbb{F}_q^\star} Pr_{x \in \mathbb{F}_q^m}[R(x) = \mu v]$. We have the following $q$-ary analogue of Corollary 7.7.

**Corollary 7.12.** *Let $L \in \mathrm{Lin}_1(\mathbb{F}_q, k, m)$ and $R : \mathbb{F}_q^k \to \mathbb{F}_q^m$ be such that $\Delta(R, L) \leqslant 1 - \frac{1}{q} - \varepsilon$. If $\mathrm{RowSpan}(L) = \langle v \rangle$ then $\mathrm{wt}(R, v) \geqslant \varepsilon$.*

Fixing the vector $v$ (from amongst $t \leqslant \frac{1}{\varepsilon}$ choices), reduces the problem to Hadamard decoding over $\mathbb{F}_q$. The Johnson bound states that at radius $1 - \frac{1}{q} - \varepsilon$, the list-size is bounded by $\frac{1}{\varepsilon^2}$. This gives the following bound.

**Lemma 7.13.** *For any $\varepsilon > 0$ we have $\ell_1^{\odot m}\big(1 - \frac{1}{q} - \varepsilon\big) \leqslant \frac{1}{\varepsilon^3}$.*

Plugging this into Lemma 7.11 completes the proof of Theorem 2.10.

Microsoft Research, Silicon Valley. Some of this work was done when the author was a postdoc at the University of Washington visiting Princeton University.

*E-mail address*: `parik@microsoft.com`

Department of Computer Science and Engineering, University of Washington. Currently visiting the Computer Science Dept., Carnegie Mellon University. Some of this work was done when the author was a member in the School of Mathematics, Institute for Advanced Study.

*E-mail address*: `venkat@cs.washington.edu`

Department of Computer Science and Engineering, University of Washington. Some of this work was done when the author was visiting Princeton University.

*E-mail address*: `prasad@cs.washington.edu`